\pdfoutput=1
\documentclass[journal]{IEEEtran}
\usepackage{xr-hyper}                   
\usepackage[utf8]{inputenc}
\usepackage{amsmath}
\usepackage{amsfonts}
\usepackage{bbding}
\usepackage{amssymb}
\usepackage{array}
\usepackage{multirow}
\usepackage{graphicx}
\usepackage{xcolor}
\usepackage[caption=false,font=footnotesize]{subfig}
\usepackage{algorithm}
\usepackage{algorithmic}
\usepackage[hidelinks]{hyperref}
\newcommand\algorithmicprocedure{\textbf{procedure}}
\newcommand{\algorithmicendprocedure}{\algorithmicend\ \algorithmicprocedure}
\makeatletter
\newcommand\PROCEDURE[3][default]{%
  \ALC@it
  \algorithmicprocedure\ \textsc{#2}(#3)%
  \ALC@com{#1}%
  \begin{ALC@prc}%
}
\newcommand\ENDPROCEDURE{%
  \end{ALC@prc}%
  \ifthenelse{\boolean{ALC@noend}}{}{%
    \ALC@it\algorithmicendprocedure
  }%
}
\newenvironment{ALC@prc}{\begin{ALC@g}}{\end{ALC@g}}
\makeatother

\usepackage{color}
\usepackage{cite}

\allowdisplaybreaks

\DeclareMathOperator{\sgn}{\bf sgn}

\DeclareMathOperator{\expt}{\mathbb{E}}

\DeclareMathOperator{\diag}{\bf diag}

\DeclareMathOperator*{\argmin}{\arg\!\min}

\usepackage{amsthm}
\usepackage{stfloats}

\newtheorem{theorem}{Theorem}

\newtheorem{rem}{Remark}

\newtheorem{prop}{Proposition}

\newcommand{\blue}[1]{{\color[rgb]{0,0,0}{#1}}}

\begin{document}
\title{\fontsize{22}{28}\selectfont Scalable Interference Graph Learning for Low-Latency Wi-Fi Networks using Hashing-based Evolution Strategy}

\author{
\IEEEauthorblockN{Zhouyou Gu, Jihong Park, Jinho Choi}\\
\thanks{
This work was supported in part by A$^*$STAR under its IAF-ICP (I2501E0064), in part by the IITP-ITRC grant funded by the Korean government (MSIT) (IITP-2026-RS-2023-00259991) (33\%), in part by SUTD Kickstarter Initiative (SKI 2021 06 08), and in part by the National Research Foundation, Singapore, and the Infocomm Media Development Authority under its Future Communications Research \& Development Programme. \textit{(Corresponding authors: J. Park and Z. Gu.)}
}  
\thanks{
Z. Gu and J. Park are with the Information Systems Technology and Design Pillar, Singapore University of Technology and Design, Singapore 487372 (email: \{zhouyou\_gu, jihong\_park\}@sutd.edu.sg).
}
\thanks{
J. Choi is with the School of Electrical and Mechanical Engineering,
the University of Adelaide, Adelaide, SA 5005, Australia
(email: \{jinho.choi\}@adelaide.edu.au).
}
\thanks{Source codes are available at {https://github.com/zhouyou-gu/scneugm-wi-fi}}
}
\maketitle
\begin{abstract}
Wi-Fi 7 introduces the restricted target wake time (RTWT) mechanism, which is vital for Industrial IoT (IIoT) applications requiring periodic, reliable, and low-latency communication. RTWT enables deterministic channel access by assigning scheduled transmission slots to stations (STAs), minimizing contention and interference. However, determining efficient RTWT slot assignments remains challenging in dense networks, where conventional interference graph-based models lack flexibility and scalability. To overcome this, we propose a scalable interference graph learning (IGL) framework that learns optimal interference graph representations for graph coloring-based RTWT scheduling. The IGL leverages an evolution strategy (ES) to train a neural network (NN) using a single network-wide reward, avoiding costly edge-wise feedback. Furthermore, a deep hashing function (DHF) groups interfering STAs, limiting training and inference to relevant subsets and greatly reducing complexity. Simulation results demonstrate that the proposed IGL improves slot efficiency by up to 25\%, reduces packet losses by up to 30\% in dynamic environments. Thanks to DHF, it also reduces the training and inference time of IGL by 4 and 8 times, respectively, and the online slot assignment time by 3 times in large networks.
\end{abstract}
\begin{IEEEkeywords} 
Interference graph, Wi-Fi, graph construction, Wi-Fi restricted target wake time.
\end{IEEEkeywords}

\section{Introduction}
\begin{figure}[!t]
\centering
\includegraphics[scale=0.6]{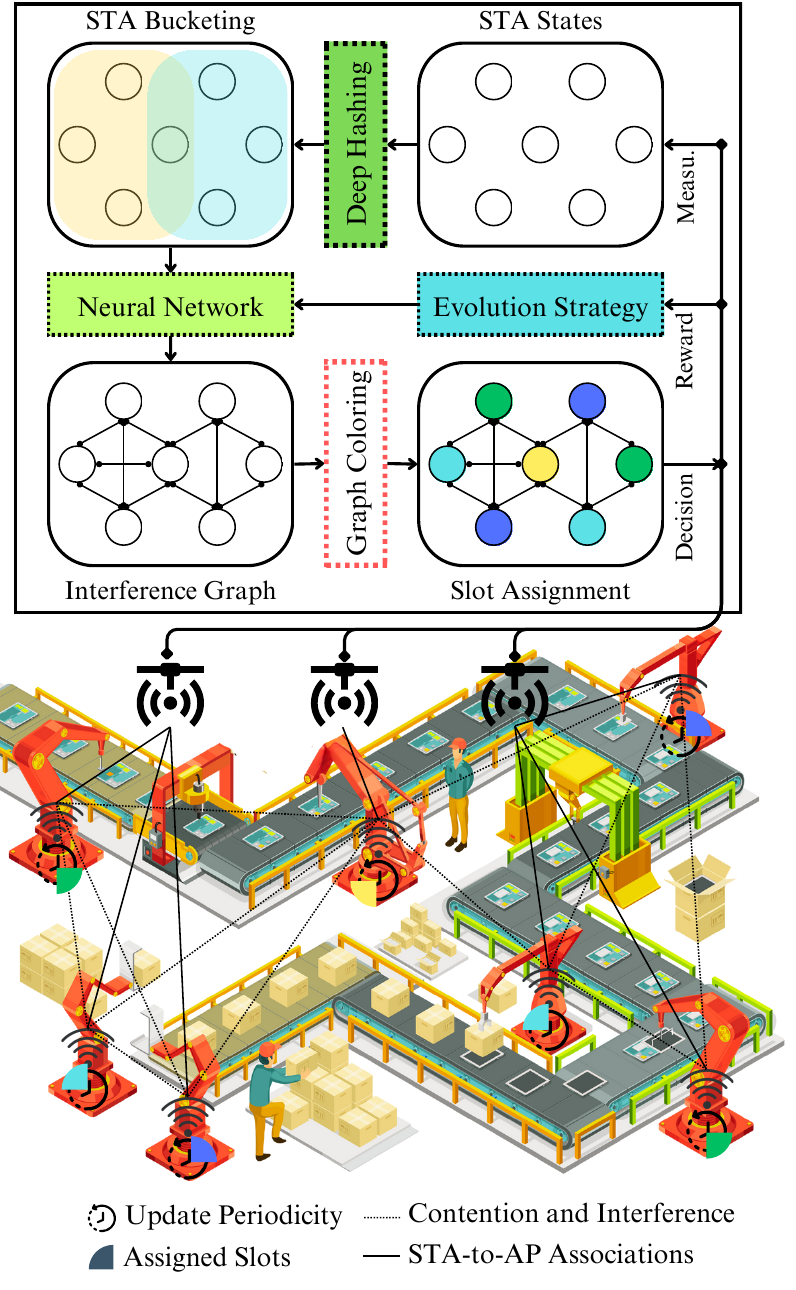}
\vspace{-0.2cm}
\caption{Illustration of a Wi-Fi network for IIoT applications with the proposed interference graph learning scheme using the hashing-based evolution strategy.}
\label{fig:contention_and_interference}
\vspace{-0.25cm}
\end{figure}

With ever-increasing demands for wireless connectivity, Wi-Fi 802.11 networks have become a vital component of modern communication infrastructure \cite{deng2020ieeeb,garcia-rodriguez2021ieee,galati-giordano2024whata}.
The evolution of Wi-Fi technology is now driven by demands for connectivity in emerging Industrial Internet of Things (IIoT) applications, including automated guided vehicles, factory automation, and tactile controls \cite{nasrallah2018ultralow,cavalcanti2022wifi,zanbouri2024comprehensive}.
In smart-factory IIoT deployments, hundreds or even thousands of sensors monitor machine health, environmental conditions, and inventory, and must report to a central controller with high timeliness and reliability to support real-time monitoring and control.
Typical IIoT networks therefore employ multiple stations (STAs) as sensors that periodically upload status updates to access points (APs) over Wi-Fi uplink channels \cite{karamyshev2024enabling}, as shown in Fig. \ref{fig:contention_and_interference}.
Since delayed or missing status updates can cause inaccurate control decisions and potential accidents \cite{3gpp2018service}, these transmissions require stringent quality-of-service (QoS), particularly low latency and high reliability.
For example, missing temperature or pressure updates in chemical processes can delay corrective actions and escalate into hazardous events, including explosions \cite{zhao2024deep}; similarly, delayed or lost perception updates in automated guided vehicles can reduce obstacle awareness and increase collision and equipment-damage risks \cite{bostelman2014methods}.

In most 802.11 families \cite{80211ieee2024}, Wi-Fi devices operate using the carrier-sense multiple access with collision avoidance (CSMA/CA) scheme for channel access, where STAs contend for transmission opportunities \cite{bianchi2000performance}.
Specifically, to avoid packet collisions with other STAs, an STA waits until the channel is free by sensing ongoing transmissions from other STAs and then backs off for a random time before transmitting its packet to avoid collisions with other STAs also waiting.
However, CSMA/CA can cause packet delays and losses due to the following two issues \cite{garetto2008modeling}. First, an STA experiences significant backoff time when it senses transmissions from many other STAs, i.e., it contends with numerous surrounding STAs. Second, concurrent transmissions can occur if STAs cannot sense each other’s transmissions due to the large distance between them, which causes inter-STA interference and further decoding errors at the receiving APs.
Note that neighboring STAs that trigger backoff or cause interference are referred to as the STA’s contending or hidden neighbors, respectively.
Consequently, the native CSMA/CA scheme in Wi-Fi can hardly provide the stringent QoS required by IIoT services due to contention and interference \cite{garetto2008modeling}.

Recent Wi-Fi 7 standards introduce the restricted target wake time (RTWT) mechanism \cite{nurchis2019targeta,yang2021target,haxhibeqiri2024coordinated}, which allocates exclusive channel access to STAs during specified time slots (or service periods). This approach improves network performance and STAs' QoS by eliminating contention and interference between STAs in different slots. When applying RTWT, STAs only need to synchronize slot boundaries and avoid frame transmissions across allocated slot boundaries \cite{khorov2019tutorial}.
This implies that RTWT has a more relaxed synchronization requirement than orthogonal frequency-division multiple access (OFDMA) \cite{deng2020ieeeb}, which requires symbol-level synchronization among STAs \cite{80211ieee2024}.
A straightforward approach is to assign each STA a dedicated slot, which completely eliminates contention and interference. However, this can lead to significant delays in large networks, as each STA must wait for many others to transmit. This issue can be mitigated by allowing slot sharing when STAs are unlikely to interfere with each other’s transmissions. The key challenge, therefore, is to design a slot assignment scheme that ensures reliable communication while minimizing the number of slots used.

\subsection{Related Works}\label{subsec:related_works}
Several approaches have been proposed to model contention and interference and to design slot assignment schemes for Wi-Fi 7 IIoT networks. We briefly discuss a few key methods.

\subsubsection{Markov Model Approach}
The CSMA/CA process in Wi-Fi STAs can be effectively modeled using Markov chains \cite{bianchi2000performance,garetto2008modeling}, capturing the impact of contention and interference from neighboring STAs.
QoS metrics for STAs, such as channel utilization \cite{chang2018traffic} and energy efficiency \cite{kai2019energy}, can be derived from the steady states of these models. However, these Markov models \cite{bianchi2000performance,garetto2008modeling} are non-linear and lack closed-form expressions for various STA transmission configurations \cite{nardelli2012closedform}, making them challenging to apply in optimizing slot assignments.
Moreover, applying these models requires prior measurements of all contending and hidden STA pairs, resulting in significant network overhead. These limitations reduce the practicality of Markov models in Wi-Fi IIoT networks.

\subsubsection{Neural Network Model Approach}
In Wi-Fi networks, neural network (NN) models can be flexibly trained to make control decisions based on STAs' QoS and network data, without requiring explicit contention and interference models \cite{szott2022wifi}.
Classic fully connected neural networks (FNNs) have pre-determined input and output dimensions, which limits their scalability to networks of varying sizes \cite{gu2021knowledge}. Instead, graph neural networks (GNNs) \cite{shen2020graph,eisen2020optimal}, which are scalable to the graph or the network size, can be applied.
GNNs can make decisions by repeatedly aggregating features (network states) of neighboring vertices (or STAs) without being constrained by network size.
When applied to the separation of contending or interfering STAs in Wi-Fi networks, the aggregation merge the features of neighboring STAs, confusing the differences in interference impacts among individual STA pairs.
This makes GNNs ineffective to separate the exact STA pairs that are heavily contending or interfering \cite{xu2018how,loukas2019what} by assigning orthogonal slots.
Additionally, since the minimum number of slots is unknown, it is challenging to design a GNN architecture capable of directly determining each STA's slot \cite{lemos2019graph,schuetz2022graph}.

\subsubsection{Human Intuition-Defined Interference Graph Approach}
STA-pairwise contention and interference information can be represented as edges between STAs, forming a graph with STAs as vertices. Slot assignment tasks can then be modeled as graph theory problems \cite{subramanian2008minimum,mishra2005weighted,chen2022energy}, such as graph cut \cite{subramanian2008minimum} or graph coloring \cite{mishra2005weighted,chen2022energy}. These tasks divide STAs into different slots (either in time or frequency) by disconnecting edges between contending or interfering STA pairs.
However, these methods construct graphs---for instance, by assigning edge values between STAs---based on fixed rules predefined according to human intuition about how contention and interference affect STAs' QoS. As a result, these graph construction methods lack flexibility and cannot be optimized to meet specific STA QoS requirements.

\subsubsection{Interference Graph Learning Approach}
NNs can be employed to generate graphs, retaining the graph structure as well as the flexibility to be trained \cite{guo2023systematic,chen2025graph}.
For example, in \cite{wang2024empowering} and \cite{zhao2025generative}, the policy gradient (PG) \cite{sutton1999policy} is to train NNs that output graphs representing activated devices and transmission links in an integrated sensing and communication system and a cyber-physical power system, respectively, where devices and links are represented as vertices and edges, respectively.
In Wi-Fi networks, the authors of \cite{gu2024graph} apply deterministic policy gradient (DPG) \cite{lillicrap2019continuous} to NNs that output an interference graph model representing STAs as vertices, with contention and interference impacts represented as weighted edges. The graph is cut based on edge weights to partition contending or interfering STAs into a fixed number of slots \cite{gu2024graph}.
However, challenges remain in minimizing slot usage and ensuring reliability in Wi-Fi IIoT networks.
Moreover, the above work \cite{gu2024graph} studies the Wi-Fi networks with a limited number of devices, e.g., up to tens of devices. Note that the number of possible device pairs grows quadratically when the number of devices increases. Thus, how to design an interference graph learning (IGL) scheme that models the edges in these massive pairs in large Wi-Fi networks requires future research.

\subsubsection*{Summary of IGL Challenges}
In large networks, each STA pair's edge decision is barely correlated with overall network performance because many other edges are influencing the performance. Therefore, PG and DPG can hardly return a performance-improving gradient direction, since they need to back-propagate each edge decision's impact on the network performance to the NN parameters. 
By contrast, evolution strategies (ESs) \cite{wierstra2014natural,salimans2017evolution} directly perturb NN parameters and evaluate NN update direction according to the perturbed parameters' performance. ESs serve as efficient black-box alternatives to the gradient-based method, especially when the credits on how each edge influences the overall performance cannot be explicitly assigned \cite{majid2024deep}. Nevertheless, how to use ESs in the IGL for large Wi-Fi networks needs further study.

Furthermore, learning interference graphs in large-scale wireless networks is computationally intensive due to the vast STA pairs, resulting in prolonged training and inference time. This delay leads to outdated interference representations that fail to capture dynamic network environments. To mitigate this, the IGL can focus on processing only the pairs of devices that are likely to contend or interfere, setting other edges to null. Given that interference is caused by the geometric proximity of devices, integrating locality-sensitive hashing (LSH) \cite{luo2023survey} can efficiently identify nearby interfering neighbors, thereby accelerating the IGL process. However, the application of LSH within the IGL remains open.

\subsection{Contributions}
This paper studies the scalable IGL approach to assign RTWT slots in Wi-Fi 7 networks, e.g., serving periodic IIoT status updates.
Our objective is to determine the optimal slot assignments that separate highly contending and interfering STAs, minimizing the number of slots required in each period while ensuring transmission reliability.
The network state information used for slot assignments includes path losses from STAs to their nearby APs and the locations of the APs, requiring no additional measurement overheads between STAs.
We model the Wi-Fi network as a binary-directed graph, where STAs are represented as vertices and the contention and interference impacts between STAs are represented as binary edges.
The coloring scheme of this graph is mapped to the RTWT slot assignment, i.e., STAs are assigned to the same slot if they share the same color, and the chromatic number of the graph is the total number of slots (colors).
We formulate the IGL task as training an IGL NN to generate the optimal interference graph, i.e., defining the edges connecting STAs, whose coloring scheme minimizes the number of slots while ensuring STAs' transmission reliability.
Two fundamental challenges in the IGL are addressed in our IGL scheme for large Wi-Fi networks:
1) The credit assignment problem \cite{majid2024deep} on estimating explicit STA-pairwise feedback on a large number of edges, since the graph's performance is measured STA-wise (STAs' reliability) or network-wise (the number of slots);
2) The high computational complexity in the IGL when all STA pairs are processed in training and inference.
To avoid acquiring STA-pairwise feedback, an ES algorithm \cite{wierstra2014natural,salimans2017evolution} is applied to directly adjust the NN parameters without evaluating individual edges.
To reduce the complexity, a deep hashing function (DHF) \cite{luo2023survey} groups likely contending and hidden STA pairs, reducing the processed pairs in the IGL.

This paper’s contributions are listed as follows.

\begin{itemize}
\item 
To the best of our knowledge, we are the first to formulate the slot minimization problem in the RTWT slot assignment as the IGL task.
Unlike GNN models \cite{shen2020graph,eisen2020optimal}, the IGL preserves the STA-pairwise interference information on the graph edges, indicating whether each STA pair will be separated or not.
Moreover, unlike human intuition-based graph modeling \cite{subramanian2008minimum,mishra2005weighted,chen2022energy} that constructs edges using fixed rules, the proposed scheme flexibly trains NNs to construct optimal graphs for specific STA QoS requirements.
Further, unlike the previous IGL in Wi-Fi \cite{gu2024graph} that is only applicable for a fixed number of slots, the proposed one minimizes the number of slots needed to satisfy STA QoS.

\item 
We apply the ES to train the NN to construct the optimal interference graph, avoiding the edge-wise credit assignment problem.
Existing works \cite{wang2024empowering,zhao2025generative,gu2024graph} use the PG or the DPG to optimize NN parameters via back-propagation, requiring the estimation of the individual edge's gradient for every edge.
Unlike these methods \cite{wang2024empowering,zhao2025generative,gu2024graph}, the proposed ES directly correlates each NN parameter's update direction with the graph's performance, regardless of the number of edges.
The NN trained by the ES generates graphs with $4$-$10$ times higher reward signals (indicating fewer slots than human intuition-based graphs while maintaining reliable transmissions) than those by the PG/DPG-trained NNs.

\item 
Also, we apply the DHF to reduce the IGL complexity by focusing on the training and inference within contending or hidden STA pairs in Wi-Fi networks.
To identify contending and hidden pairs, the DHF embeds each STA state into a hash code, where the bits of two STAs' codes are likely equal if they form a contending or hidden pair.
By querying random bits in the STAs' codes, two STAs are grouped if their bits are likely equal, i.e., they are likely contending or hidden pairs. In large networks with $1000$ STAs, the DHF reduces the training and inference time of the IGL by up to $4$ and $8$ times, respectively, and the online slot assignment time by up to $3$ times.

\item
We implement and deploy the IGL in a system-level, standard-compliant Wi-Fi simulation platform, NS-3 \cite{nsnam}, with RTWT enabled.
Simulations show that graphs returned by the ES-trained NN reduce the number of time slots by $25\%$ compared to graphs designed based on human intuition at the cost of slightly increased QoS violations, e.g., additional $1\%$ packet loss.
Further, in online scenarios with mobile STAs, the proposed IGL achieves up to $30\%$ fewer packet losses by applying the DHF to efficiently group STA pairs, enabling timely graph regeneration adaptive to network dynamics compared to the scheme processing all STA pairs non-selectively.

\end{itemize}

\subsection{Notation and Paper Organization} 
The $i$-th element of a vector $\mathbf{x}$ is denoted as $x_{i}$.
The $j$-th element of the $i$-th row of a matrix $\mathbf{X}$ is denoted as $X_{i,j}$.
We define the elements of $\mathbf{X}$ as $\mathbf{X} \triangleq [X_{i,j} \mid X_{i,j} = (\cdot)]$, where $(\cdot)$ represents the expression defining $\mathbf{X}$'s elements.
$\diag{\mathbf{X}}$ denotes the diagonal elements of $\mathbf{X}$.
$\mathbf{1}_{\{\cdot\}}$ is an indicator that equals $1$ if $\{\cdot\}$ is true and $0$ otherwise.
$\land$ is logical AND.

The rest of this paper is organized as follows.
Section \ref{sec:wifi_system_model} presents the system model. Section \ref{sec:ScNeuGM_framework} defines the IGL task for contention and interference management and explains the overall concept of the IGL framework. 
Sections \ref{sec:ggm} and \ref{sec:dhf} present the ES and the DHF in the IGL, respectively. 
Finally, Section \ref{sec:simulation_results} shows the simulations evaluating the proposed methods, and Section \ref{sec:conclusion} concludes this work.

\section{System Model}\label{sec:wifi_system_model}
This section presents the system model of the Wi-Fi network, including background on the RTWT mechanism.

\subsection{Wi-Fi Network with Multi-AP
Coordination}\label{subsec:wifi_net_configuration}
\begin{figure}[!t]
\centering
\includegraphics[scale=0.9]{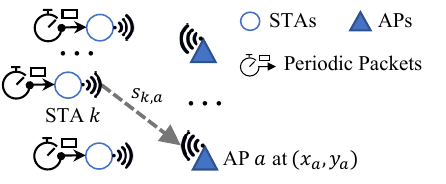}
\vspace{-0.2cm}
\caption{System model of the Wi-Fi IIoT network.}
\label{fig:wi-fi_network_system_model}
\vspace{-0.25cm}
\end{figure}

In the network, STAs are assumed to be sensors or monitors continuously collecting data to track critical information in IIoT applications, such as manufacturing systems, and transmitting the collected data samples to the APs via the uplink, as shown in Fig. \ref{fig:contention_and_interference}.
We consider a Wi-Fi network consisting of $K$ STAs and $A$ APs, as illustrated in Fig. \ref{fig:wi-fi_network_system_model}.
The APs are located in a two-dimensional space, and each AP's location is known in advance as $\mathrm{loc}_a \triangleq (x_a, y_a)$, where $x_a$ and $y_a$ are the coordinates of AP $a$ along the $x$ and $y$ axes, respectively, with $a \in \{ 1, \dots, A\}$.
All STAs and APs are assumed to operate on the same channel with a bandwidth of $B$ in Hertz (Hz). The transmission power of the STAs is $\mathbf{P}_\mathrm{0}$ in milliwatts (mW), and the noise power spectral density is $\mathbb{N}_\mathrm{0}$ in mW/Hz.

The APs are coordinated by a single controller that manages the STAs' connections. The controller runs at an edge server that has real-time communication with all APs, via which it receives network measurements and sends control decisions from and to APs. 
The coordinated APs form an extended service set \cite{80211ieee2024} where STAs can roam between APs without resending re-association requests to APs.
Each STA is associated with the AP that has the minimum path loss to the STA, and the associated AP receives and collects the STA's transmitted packets.
We denote the path loss from STA $k$ to AP $a$ as $s_{k,a}$ in dB, $\forall k, a$ and the associated AP of STA $k$ as $\hat{a}_k$, i.e.,
\begin{equation}\label{eq:associated_ap_id}
\begin{aligned}
\hat{a}_k \triangleq \argmin_{a \in \{1,\ldots, A\}} s_{k,a}  .
\end{aligned}
\end{equation}
Each AP keeps measuring its path losses to the STAs based on the STAs' packets detected and received at the AP. 
The measured path losses will be updated to the controller that decides which AP should be associated with the STA and process the STA's packets.
We assume that the same data-collection mechanism is used in all STAs over time and that STAs encapsulate the collected data sample into a packet with a constant format, i.e., each STA has the same packet arrival process with a packet size of $L$ bits.

The controller will notify each STA of the AP that it is associated with by instructing the APs to send a notification message. The message contains the path loss from the STA to the associated AP whenever the association changes or the path loss changes significantly. Then, each STA can adjust the modulation and coding scheme (MCS) of its packet transmissions according to the new path loss.
Specifically, let $d_k$ (in seconds) denote the transmission duration of an $L$-bit packet for STA $k$, which depends on the MCS.
Each STA's MCS is configured to be the highest one for which the decoding error probability without interference, $\epsilon_k$, is below a threshold, say $\epsilon_{\max} = 10^{-5}$.
The decoding error probability $\epsilon_k$ for short packets can be approximated as \cite{yang2014quasi}
\begin{equation}\label{eq:tx_error_for_mcs_wi-fi}
\begin{aligned}
\epsilon_k \approx f_Q\Bigg(\frac{  -L\ln{2}+ {d_k B}\ln(1+\phi_k)}{\sqrt{d_k B V}}\Bigg),\ \forall k ,
\end{aligned}
\end{equation}
where $\phi_k = \frac{1}{s_{k,\hat{a}_k}} \frac{\mathbf{P}_\mathrm{0}}{\mathbb{N}_\mathrm{0}B}$ is the signal-to-noise-ratio (SNR) at STA $k$ without interference ($s_{k,\hat{a}_k}$ here is converted to the decimal scale).
In \eqref{eq:tx_error_for_mcs_wi-fi}, $f_Q$ is the tail distribution of the standard normal distribution, and $V$ is the channel dispersion as $V = 1- \frac{1}{(1+\phi_k)^2}$ \cite{yang2014quasi} in \eqref{eq:tx_error_for_mcs_wi-fi}.
Note that we assume that $\mathbf{P}_\mathrm{0}$, $\mathbb{N}_\mathrm{0}$, $B$, and $L$ in \eqref{eq:tx_error_for_mcs_wi-fi} are constants in the network. Thus, for each STA $k$, the packet duration $d_k$ only depends on the path loss to STAs' associated APs, $s_{k,\hat{a}_k}$, $\forall k$. Throughout the paper, we assume that $d_k$ is bounded and less than the slot duration.

\subsection{CSMA/CA with RTWT}
\begin{figure}[!t]
\centering
\includegraphics[scale=1.2]{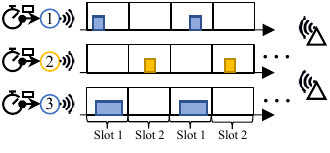}
\vspace{-0.2cm}
\caption{Illustration of the slot assignments for $K=3$ STAs with $Z=2$ slots, where  STA $1$ and $3$ are assigned in the same slot, e.g., $z_1=z_3=1$, and STA $2$ is assigned with a different one, e.g., $z_2=2$.}
\label{fig:slot_assignment}
\vspace{-0.25cm}
\end{figure}

We assume that all STAs and APs are synchronized in time, and the time is divided into slots indexed by $t \in \{0,1,\ldots\}$. The duration of each slot is denoted as $\Delta_\mathrm{0}$ seconds. For RTWT service, denote by $Z$, which is a positive integer, the periodicity for all STAs, while $z_k \in\{1,\ldots, Z\}$ represents the offset of STA $k$ its RTWT service.
The periodicity $Z$ directly affects the freshness of status updates from STAs. A larger $Z$ increases the delay between updates, while a smaller $Z$ causes more STAs to share the same slots, leading to higher contention and interference, which can degrade transmission reliability.
Here, we assume that the slot duration $\Delta_\mathrm{0}$ is fixed while $Z$ can be adjusted to increase or decrease the total RTWT period duration $Z \Delta_\mathrm{0}$ to balance the update freshness and contention/interference levels.
Specifically, STA $k \in \{1,\ldots,K\}$ is scheduled to transmit exclusively during the periodic time slots $t = z_k, z_k + Z, z_k + 2Z, \dots$. Note that multiple STAs can transmit in the same time slot if their transmission offsets are identical.
For example, as shown in Fig.~\ref{fig:slot_assignment}, 
suppose there are three STAs ($K = 3$) and that $Z = 2$. In this illustrated example, contention and interference between STAs 2 and 1, as well as between STAs 2 and 3, are eliminated, while contention and interference may still occur between STAs 1 and 3.
Each STA contends for channel access using CSMA/CA \cite{80211ieee2024} during its periodically assigned slots. The AP sends a negative acknowledgment if the packet transmission fails, triggering the STA to retransmit the same packet. Retransmission repeats until the maximum number of attempts is reached, an acknowledgment is received or the assigned slot has expired. Afterward, the STA waits for the next assigned slot in the subsequent period.
Consequently, contention and interference occur only among STAs with the same offset, while such conflicts are avoided between those with different slot assignments.

\subsection{Wi-Fi Network States}\label{subsec:wifi_net_states}

This subsection describes the network state information that will be used to discuss the proposed slot assignment scheme.

\subsubsection{Measured Network States}

In this paper, Wi-Fi devices are assumed to operate with practical receiver sensitivities. Specifically, if the path loss between a STA and an AP exceeds a threshold $s_{\max}$, the AP is assumed to be unable to detect or receive transmissions from the STA, and such path loss values cannot be measured by the network. To this end, for each STA, the APs capable of detecting its signal are identified and sorted into a list based on the measured path losses as follows:
\begin{equation}\label{eq:sorted_ap_list}
\begin{aligned}
\mathcal{A}_k \triangleq \{a_k(1),a_k(2),\dots,a_k(|\mathcal{A}_k|)\},\ \forall k \ ,
\end{aligned}
\end{equation}
where $s_{k,a} \leq s_{\max}, \ \forall a \in \mathcal{A}_k$, and $|\mathcal{A}_k|$ denotes the number of APs capable of detecting STA $k$'s signal. The APs in \eqref{eq:sorted_ap_list} are sorted in ascending order of path loss from STA $k$, such that $s_{k,a_k(i)} \leq s_{k,a_k(j)}, \ \forall i < j$.
Clearly, $a_k(1) = \hat{a}_k$, as $a_k(1)$ represents the AP with the minimum path loss to STA $k$, and STA $k$ is associated with this AP, as mentioned in Section \ref{subsec:wifi_net_configuration}.
We denote STA $k$'s path loss to an AP in the above list, along with the location of the AP, as a state vector:
\begin{equation}
\begin{aligned}
\mathbf{s}_k(i) \triangleq [s_{k,a_k(i)},x_{a_k(i)},y_{a_k(i)}]^{\rm T}  ,\ i=1,\dots, |\mathcal{A}_k|  , \ \forall k  .
\end{aligned}
\end{equation}
The sequence of sorted STA $k$'s state vectors is written as
\begin{equation}
\begin{aligned}
\mathcal{S}_k \triangleq \{\mathbf{s}_k(1),\mathbf{s}_k(2),\dots,\mathbf{s}_k(|\mathcal{A}_k|)\}  ,\ \forall k ,
\end{aligned}
\end{equation}
Finally, the collection of all such sequences is defined as
\begin{equation}
\begin{aligned}
\hat{\mathcal{S}} \triangleq \{\mathcal{S}_1,\dots,\mathcal{S}_K\},
\end{aligned}
\end{equation}
which is referred to as the collection of network states, with $\mathcal{S}_k$ representing the state of STA $k$. Note that the size of each STA's state, $\mathcal{S}_k$, may vary since STAs can have different numbers of APs capable of detecting their signals, depending on the locations of the STAs and APs.
Here, $\hat{\mathcal{S}}$ represents the measured network states and is maintained at the edge controller as the input to the proposed slot assignment methods. 

\subsubsection{Unmeasured Network States}
Whether a pair of STAs is a contending or hidden pair significantly impacts their channel access and transmission reliability \cite{garetto2008modeling}. Contending and hidden STAs depend on the path losses between STA pairs, specifically on whether two STAs can detect each other. Measuring the path losses for all $K$ STAs requires approximately $\mathcal{O}(K^2)$ measurements for all STA pairs, which introduces significant overhead to the network. Therefore, we assume that contending and hidden STA relationships are not measured during the online deployment of the proposed methods but are instead used as training information in offline simulations.
Specifically, contending and hidden STA pairs, $\forall i,j$, are represented by binary indicators as
\begin{equation}\label{eq:contending_and_hidden_sta_indicators}
\begin{aligned}
O^{\mathrm{C}}_{i,j}\triangleq \mathbf{1}_{\{g(i,j)\leq s_{\max}\}};O^{\mathrm{H}}_{i,j}\triangleq \mathbf{1}_{\{g(i,j)> s_{\max} \}}\land \mathbf{1}_{ \{s_{i,\hat{a}_j} \leq s_{\max} \}}.
\end{aligned}
\end{equation}
We set $O^{\mathrm{C}}_{i,j}=1$ if the path loss $g(i,j)$ between STAs $i$ and $j$ is less than or equal to the threshold $s_{\max}$, indicating STA $i$ is contending with STA $j$, and set $O^{\mathrm{H}}_{i,j}=1$ if the path loss $g(i,j)$ between STAs $i$ and $j$ exceeds $s_{\max}$ and STA $i$'s path loss to STA $j$'s associated AP, $s_{i,\hat{a}_j}$, is less than or equal to $s_{\max}$, where STA $i$ interferes with STA $j$'s AP and is hidden from STA $j$.
$O^{\mathrm{C}}_{i,j}+O^{\mathrm{H}}_{i,j}$ denotes that an STA pair $(i,j)$ is either contending or hidden STAs. 
We assume that if two STAs are neither contending nor hidden STAs, their transmissions do not affect each other.

\section{The Proposed IGL Framework}
\label{sec:ScNeuGM_framework}

In this section, we first introduce the problem formulation for contention and interference management. We then transform the problem into an interference graph-constructing task, where optimizing the edge weights corresponds to finding an optimal coloring scheme, which in turn determines the slot assignment. Finally, we present the IGL framework, which trains a neural network to generate the optimal graph structure while addressing the challenges posed by large-scale networks.

\subsection{Problem Formulation}\label{subsec:problem_formulation}

To improve QoS and reduce packet loss rates, we can assign the slot choices of STAs $\mathbf{z} \triangleq [z_1, \dots, z_K]$ in the period, where $z_k \in \{1, \dots, Z\}$. Let $u_k(n)$ denote the number of packets successfully sent by STA $k$ during the $n$-th slot. Note that $u_k(n) \in \{0, 1\}$ since each STA transmits only one packet per slot, as mentioned in Section \ref{sec:wifi_system_model}. Then, the reliability of each STA is defined as the successful packet transmission rate averaged over time as
\begin{equation}\label{eq:asymptotic_average_user_throughput}
\begin{aligned}
r_k
&\triangleq \lim_{N\to \infty} \frac{1}{N} \sum_{n=1}^{N} u_k(n),\ \forall k.
\end{aligned}
\end{equation}
Note that unsuccessful transmissions can occur due to insufficient transmission time in each assigned slot (e.g., caused by contention backoffs) or due to decoding errors (e.g., caused by interference). 

Suppose that we have a QoS requirement such that each STA's reliability must exceed a threshold $\hat{r}$. Specifically, the expected value of the reliability of STA $k$, given the measured network states and slot assignment decisions, $\mathbb{E}[r_k|\hat{\mathcal{S}},\mathbf{z}]$, must be greater than or equal to $\hat{r}$, which becomes the constraint. The objective is to minimize the number of slots, $Z$, in the period or, equivalently, the transmission periodicity.
Minimizing $Z$ is essential for preserving update freshness. A smaller $Z$ enables more frequent status reporting, whereas a larger one increases the risk of stale information, potentially leading to suboptimal control actions and safety-critical failures. In IIoT applications that require real-time monitoring and control, optimizing $Z$ is therefore fundamental to maintaining timely data collection and delivery from STAs \cite{nasrallah2018ultralow,cavalcanti2022wifi,zanbouri2024comprehensive}. 
Finally, the optimization problem is given by
\begin{equation}\label{eq:prob:cni_management_problem}
\begin{aligned}
&\min_{\mathbf{z},Z}\ Z \cr 
&\text{s.t. } \ \expt[r_k|\hat{\mathcal{S}},\mathbf{z}] \geq \hat{r} , \ z_k \in \{1,\dots,Z\} , \  \forall k .
\end{aligned}
\end{equation}
However, the straightforward formulation in \eqref{eq:prob:cni_management_problem} is difficult to solve due to the lack of a clear structure for deriving the slot assignments and the minimum number of slots from the network states. To address this challenge, the problem is reformulated as a graph-constructing task, which will be discussed in the following section.


\subsection{Problem Transformation to Graph-Construction Task}
In Wi-Fi networks, contention and interference are pairwise interactions between STAs. To effectively represent these STA-pairwise interactions, we model the network as a binary-directed graph, $\mathcal{G} = (\mathcal{V}, \mathcal{E})$, where the STAs are vertices $\mathcal{V}$ and each directed edge $(i,j) \in \mathcal{E}$ implies that STA $i$'s transmissions will cause a QoS violation for STA $j$'s transmissions if they are in the same slot. 
Specifically, the QoS violations, i.e., packet losses, happen 1) when one STA contends the channel with the other STA as they are close to each other, causing a significant backoff time and delaying the packet transmission until no sufficient time remaining in the slot; 2) when one STA is hidden from the other STA and is close to the other STA's receiving AP, causing interference at the receiving AP.
The adjacency matrix $\mathbf{E}$ of $\mathcal{G}$ is defined with binary edges as
\begin{equation}\label{eq:defi:graph_adjacency_matrix}
\begin{aligned}
\mathbf{E} \triangleq \big[E_{i,j}\big|E_{i,j}\in\{0,1\},\ \forall i\neq j;\ E_{k,k}=0,\ \forall k \big] .
\end{aligned}
\end{equation}
Graph coloring is employed to assign colors to vertices (STAs) such that no two adjacent vertices share the same color. Each color represents a distinct group of STAs that can transmit simultaneously without causing QoS violations, i.e., they can be assigned to the same slot. If an edge $(i,j)$ exists (i.e., $E_{i,j} = 1$), STAs $i$ and $j$ must be assigned different colors to avoid contention and interference.

\begin{rem}
  The graph can be also be constructed as a weighted graph, where the edge weights represent the severity of contention and interference between STAs.
  In this case, this STA is more likely to be separated from its neighboring STAs when the sum of its edge weights is large (implying a significant aggregated interference or contention). We can achieve the above separation rules by using weighted graph coloring, where the sum of edge weights from neighbors with the same color is required to be less than an interference threshold. For simplicity, we assume that the edge weights can only take $0$ or $1$ and the interference threshold is $1$, considering the binary interference effect on QoSs and ignoring the aggregated small interference case, while the proposed methods can be straightforwardly extended to weighted graphs \cite{gu2024graph}, utilizing weighted graph coloring algorithms \cite{west2001introduction}.
\end{rem}

We denote the coloring scheme of the binary graph with adjacency matrix $\mathbf{E}$ as
\begin{equation}\label{eq:defi:graph_coloring}
\begin{aligned}
\{\mathbf{z}, Z\} = \chi(\mathbf{E}) ,
\end{aligned}
\end{equation}
where $\mathbf{z}=[z_1,\dots,z_K]^{\rm T}$ is decided as the proper color choices for STAs $1,\dots,K$ that satisfy the edge constraints on the graph, and $Z$ now represents the minimum number of colors required to color the graph (i.e., the graph's chromatic number), where $z_k\in\{1,\dots,Z\}$ $\forall k$. The coloring scheme is mapped to the slot assignments as follows.

\begin{prop}\label{prop:existence_of_optimal_binary_graph}
There exists an adjacency matrix $\mathbf{E}^*$ of the graph whose coloring scheme, $\{\mathbf{z}^*, Z^*\} = \chi(\mathbf{E}^*)$, solves the contention and interference management problem in \eqref{eq:prob:cni_management_problem}. In other words, $\mathbf{z}^*$ and $Z^*$ represent the optimal slot assignments in \eqref{eq:prob:cni_management_problem}, minimizing the number of slots while satisfying the STAs' reliability constraints.
\end{prop}
\begin{proof}
See Appendix A
\end{proof}

Consequently, by optimizing the graph's binary edge values and then coloring the graph, we can equivalently optimize the slot assignments in the problem \eqref{eq:prob:cni_management_problem}. Based on this fact, we construct the following graph optimization task as
\begin{equation}\label{eq:prob:cni_management_problem:gc}
\begin{aligned}
& \min_{E_{i,j}\in \{0,1\},\forall i\neq j}  \ Z \cr
&\ \textrm{s.t.} 
 \ \expt[r_k|\hat{\mathcal{S}},\mathbf{z}] \geq \hat{r} , \  \forall k , \\
&\quad\quad \{\mathbf{z},Z \}= \chi(\mathbf{E}).\\
\end{aligned}
\end{equation}
where the graph coloring in \eqref{eq:defi:graph_coloring} is embedded to decide the STAs' slots, and the edge values of the graph are optimized instead. 
Since there exists an optimal graph as
\begin{equation}
\begin{aligned}
\{\mathbf{z}^*,Z^* \}= \chi(\mathbf{E}^*),
\end{aligned}
\end{equation}
whose coloring scheme is the optimal slot assignment, as shown in Proposition \ref{prop:existence_of_optimal_binary_graph}, 
we can find the optimal slot assignments in \eqref{eq:prob:cni_management_problem} by solving the problem \eqref{eq:prob:cni_management_problem:gc} instead.
As a result, this formulation in \eqref{eq:prob:cni_management_problem:gc} transforms the slot assignment problem into an interference graph-constructing task that finds the optimal binary graph representing the impact of the contention and interference on STAs' transmissions.

\subsection{Interference Graph Learning}\label{subsec:igl_framework}
We simplify the problem by assuming that each binary edge value in the optimal graph, $E^*_{i,j}$, is a function of the states of STA $i$ and STA $j$, $\mathcal{S}_i$ and $\mathcal{S}_j$, as\footnote{
In practice, the interference relationship may be influenced by other STAs in the network, especially in dense deployments. Future work can explore more complex models that consider higher-order interactions among multiple STAs when determining edge values, e.g., by incorporating GNN architectures to generate the graph edges.
}
\begin{equation}\label{eq:assu:A_equal_to_mu_si_sj}
\begin{aligned}
E^*_{i,j} \triangleq \mu^*(\mathcal{S}_i,\mathcal{S}_j),\ \forall i\neq j ,
\end{aligned}
\end{equation}
where we refer to $\mu^*(\cdot)$ as the optimal graph-constructing function, mapping each STA pair's states to the optimal edge value between them. This assumption further transforms the slot assignment task into the problem of finding the optimal graph-constructing function $\mu^*(\cdot)$.

In practice, the optimal graph model is unknown and is difficult to derive analytically due to the complex interactions among STAs and their associated APs.
Therefore, the edge values have been heuristically designed based on human intuition on how contention and interference impact the network.
For example, Edges can be designed as $\mu^*(\mathcal{S}_i,\mathcal{S}_j)\approx \mathbf{1}_{\{\mathcal{A}_i\cap \mathcal{A}_j \neq \emptyset\}}, \forall i,j$ \cite{chen2022energy}, where $\mathcal{A}_i,\forall i$ can be computed using $\mathcal{S}_i$ as \eqref{eq:sorted_ap_list}.
However, human intuition-based designs rely on a fixed function and therefore cannot be adjusted to optimally capture the impact of contention and interference on STAs' reliability or minimize the number of slots. 
Specifically, the following factors will impact the interaction between two STAs $i$ and $j$ and differentiate how one STA's transmission affects the other STA's transmission:
\begin{itemize}
    \item $s_{i,\hat{a}_i}$: Path loss from STA $i$ to its associated AP $\hat{a}_i$, which determines the packet duration of STA $i$, i.e., the interference duration caused by STA $i$. A smaller $s_{i,\hat{a}_i}$ results in shorter interference durations from STA $i$.
    
    \item $s_{i,\hat{a}_j}$: Path loss from STA $i$ to STA $j$'s AP $\hat{a}_j$. It indicates the interference power at the AP. A smaller $s_{i,\hat{a}_j}$ indicates higher interference from STA $i$ to STA $j$'s transmissions.

    \item $s_{j,\hat{a}_j}$: Path loss from STA $j$ to its associated AP $\hat{a}_j$. It determines STA $j$'s received signal power and packet duration. A smaller $s_{j,\hat{a}_j}$ indicates a shorter packet duration and a stronger signal, making STA $j$'s transmissions more robust against contention and interference.

    \item $O^{\mathrm{C}}_{i,j}$ and $O^{\mathrm{H}}_{i,j}$: Indicators on the STA pair being contending or hidden, where packet losses are caused by backoffs at transmitting STAs or low SINRs at receiving APs, respectively. As these indicators are not measured, we predict them based on measured path losses.
\end{itemize}
These factors vary every STA pair, and how they are quantitatively related to the QoS violations is difficult to derive. Thus, it is hard to design the optimal graph-constructing function manually.
This issue motivates us to use IGL that approximates the unknown structure of $\mu^*(\cdot)$ using an NN \cite{gu2024graph}, as
\begin{equation}\label{eq:mu_approx_by_nn}
\begin{aligned}
E_{i,j}=\mu(\mathcal{S}_i,\mathcal{S}_j|\theta^\mu) \approx \mu^*(\mathcal{S}_i,\mathcal{S}_j),\ \forall i\neq j,
\end{aligned}
\end{equation}
where $\theta^\mu$ denotes $\mu(\cdot)$'s parameters that can be flexibly trained using machine learning (ML). We refer to $\mu(\cdot|\theta^\mu)$ as the IGL NN for Wi-Fi contention and interference management.


\begin{figure}[!t]
\centering
\includegraphics[scale=0.9]{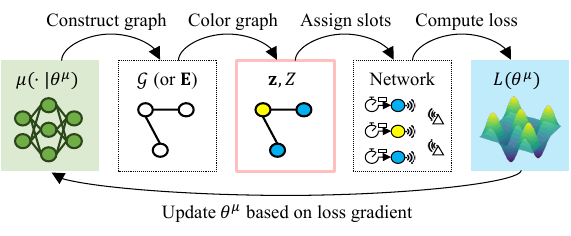}
\vspace{-0.2cm}
\caption{Iterative training of the IGL NN $\mu(\cdot|\theta^\mu)$ for the contention and interference management problem in the Wi-Fi network.}
\label{fig:ggm_flow}
\vspace{-0.25cm}
\end{figure}

Fig. \ref{fig:ggm_flow} illustrates the iterative training of the IGL NN for managing contention and interference. The process begins with the IGL NN, $\mu(\cdot|\theta^\mu)$, computing the edge values $\mathbf{E}$ between STA pairs to generate the binary graph model $\mathcal{G}$. This graph is then colored to obtain the slot assignment decision, $\mathbf{z}$ and $Z$. The network evaluates the performance of the decision and computes the loss, $L(\theta^\mu)$, on the NN parameters $\theta^\mu$. Finally, the NN parameters $\theta^\mu$ are updated based on the loss value using the ML algorithm, and the workflow repeats until the NN is optimized.
Note that graph coloring is an NP-hard task. To perform graph coloring efficiently with low complexity, we employ a greedy coloring scheme \cite{west2001introduction} that prioritizes STAs with higher degrees. Meanwhile, designing coloring algorithms is beyond the scope of this work.

\subsection{Challenges of Scaling IGL in Large Wi-Fi Networks}
When scaling the IGL to a large network with a significant number of Wi-Fi devices, several issues arise due to the massive number of STA pairs, degrading IGL performance. The proposed IGL framework, described in the following sections, addresses these issues as follows.

\subsubsection{Implicit Credit Assignment on Edge Values}


In the iteration of the IGL training, the NN parameters $\theta^\mu$ are updated in the direction of minimizing the loss $L(\theta^\mu)$, i.e., the gradient of the loss w.r.t. the parameters, $-\nabla_{\theta^\mu}L(\theta^\mu)$. Typically, this gradient is estimated using the chain rule \cite{wang2024empowering,gu2024graph}, which decomposes into the gradient of the loss w.r.t. each edge value and the gradient of the edge value w.r.t. the NN parameters as
\begin{align}\label{eq:chain_rule_for_graph}
-\nabla_{\theta^\mu}L(\theta^\mu)  = - \nabla_{\mathbf{E}} L(\theta^\mu)\nabla_{\mathbf{\theta^\mu}} \mathbf{E} = - \sum_{i \neq j}\frac{\partial L(\theta^\mu)}{\partial E_{i,j}}\nabla_{\mathbf{\theta^\mu}} E_{i,j}.
\end{align}
However, the partial derivative of the loss w.r.t. each edge value, $\partial L(\theta^\mu)/\partial E_{i,j}$, can hardly be explicitly estimated. This is because the graph's performance is measured either STA-wise (e.g., each STA's reliability) or network-wide (e.g., the number of slots used), and there is no edge-wise (or STA-pairwise) feedback. Meanwhile, the graph's performance is highly sensitive to individual edge values. For example, if an STA pair is heavily contending or interfering with each other but the IGL NN fails to set $1$ on the edge between them, the reliability constraints of the STAs could be violated. Conversely, if the IGL NN sets $1$ on edges between non-contending and non-hidden STAs, the chromatic number of the graph will increase, preventing the minimization of the number of slots. 
Since many edges in large networks can cause sensitive performance fluctuations, assigning ``credit" on how each edge impacts the network is difficult \cite{majid2024deep}. Thus, edge gradient estimation in \eqref{eq:chain_rule_for_graph} fails to provide efficient update directions for the IGL NN parameters when there is no explicit STA-pairwise feedback to individual edges.
For example, in a network with many STAs, changing a single edge may not lead to a noticeable change in the overall loss, while the variation in loss is also not directly attributable to specific edges. Therefore, we cannot approximate the gradient of the loss with respect to the parameters, which leads to the failure the DPG/PG methods, as shown in the simulation results.
To address this issue, the IGL uses the ES algorithm, an edge-feedback-free method, for estimating the update directions of the IGL NN in Section \ref{sec:ggm}.

\subsubsection{Quadratic Complexity in Network Size}
The number of possible STA pairs in the network grows quadratically with the number of STAs. For instance, in a network with one thousand STAs, there would be one million STA pairs. Simulating the interactions and computing the edges between all pairs will cause a significant computational load in large networks.
The computation can be intuitively accelerated by computing edges only between contending or hidden STA pairs, which is formally stated as
\begin{theorem}\label{theorem:existence_of_optimal_binary_graph_based_on_CH}
For any optimal graph $\mathcal{G}^*$, the graph $\mathcal{G}'$ with edges defined as $E'_{i,j} = E^*_{i,j} \land (O^{\mathrm{C}}_{i,j} + O^{\mathrm{H}}_{i,j})$ is also optimal. Here, $O^{\mathrm{C}}_{i,j}+O^{\mathrm{H}}_{i,j}$ indicates whether STAs $i$ and $j$ are contending or hidden, as defined in Section \ref{sec:wifi_system_model}.
\end{theorem}
\begin{proof}
See Appendix B.
\end{proof}
The above statement implies that any optimal graph remains optimal if edges between non-contending and non-hidden STA pairs are removed. It further suggests that the IGL only needs to consider the edges for contending and hidden pairs, while the remaining edges can be ignored and set to $0$. However, since the indicators of contending and hidden pairs, $O^{\mathrm{C}}_{i,j}$ and $O^{\mathrm{H}}_{i,j}$ $\forall i,j$, are not measured in online deployment, it is unknown which STA pairs require the IGL operations.
To address this issue, the IGL uses a low-complexity method based on the DHF to efficiently select the contending and hidden STA pairs in Section \ref{sec:dhf}.

\section{Parameter Space Optimization of IGL NN Using Evolution Strategy in IGL}\label{sec:ggm}
This section first presents the structure of the IGL NN, followed by the ES algorithm training the NN in the IGL.

\subsection{Design of IGL NN Structure}\label{subsec:ggm_structure}
\begin{figure}[!t]
\centering
\includegraphics[scale=0.9]{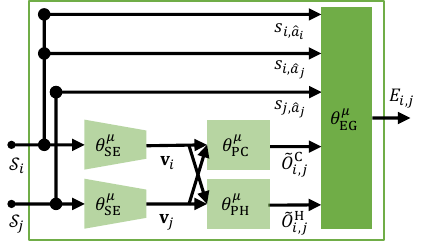}
\caption{The structure of the IGL NN $\mu(\cdot|\theta^\mu)$. Here, the state embedding with parameters $\theta^{\mu}_{\mathrm{SE}}$, the predictors with parameters $\theta^{\mu}_{\mathrm{PC}}$ and $\theta^{\mu}_{\mathrm{PH}}$ are trained using unsupervised/supervised learning in Appendix C, and the edge generator with parameters $\theta^{\mu}_{\mathrm{EG}}$ are trained using the ES in Algorithm \ref{alg:es-ggm}.
Note that $s_{i,\hat{a}_i}$, $s_{i,\hat{a}_j}$, and $s_{j,\hat{a}_j}$ are extracted from the state sequences $\mathcal{S}_i$ and $\mathcal{S}_j$.
}
\label{fig:ggm_structure}
\vspace{-0.25cm}
\end{figure}

The structure of the IGL NN $\mu(\cdot|\theta^\mu)$ is shown in Fig.~\ref{fig:ggm_structure}, consisting of 1) a state embedding that maps the STA state sequences in \eqref{eq:sorted_ap_list} into fixed-dimensional embedding vectors, 2) predictors that infer contention and interference between STA pairs, and 3) an edge generator that outputs the binary edge value. 
The design of each component is explained as follows.

\subsubsection{State Embedding}
We design the state embedding NN (SENN) as an autoencoder \cite{hinton2006reducing} that embeds the STA state sequences into a vector. 
To embed the sequence, multi-layer long short-term memory (LSTM) NNs are employed to construct the encoder \cite{sutskever2014sequencea}. Specifically, the $\mathrm{SENN}(\cdot|\theta^{\mu}_{\mathrm{SE}})$, with parameters $\theta^{\mu}_{\mathrm{SE}}$, computes each STA's embedding vector as
\begin{equation}\label{eq:tokenizing_state}
\begin{aligned}
\mathbf{v}_k = \mathrm{SENN}(\mathcal{S}_k|\theta^{\mu}_{\mathrm{SE}}), \ \forall k ,
\end{aligned}
\end{equation}
where internally LSTMs generate the embedding vector as
\begin{equation}
\begin{aligned}
\mathbf{h}_a^{(l)}, \mathbf{c}_a^{(l)} = \mathrm{LSTM}^{(l)}(\mathbf{h}_a^{(l-1)}, \mathbf{h}_{a-1}^{(l)}, \mathbf{c}_{a-1}^{(l)}|\theta^{\mu}_{\mathrm{SE}}), \ \forall l, a .
\end{aligned}
\end{equation}
where $l=1,\dots,\xi$ is the layer index and $a=1,\dots,|\mathcal{A}_k|$ is the position index in the sequence $\mathcal{S}_k$. Here, $\mathbf{h}_a^{(l)}$ and $\mathbf{c}_a^{(l)}$ are the hidden and cell vectors in the $(l,a)$-th LSTM block, $\forall l, a$.
The initial hidden state is set as $\mathbf{h}_0^{(l)}=\mathbf{0}$ for $\forall l$. The input to the first layer, $\mathbf{h}_a^{(0)} = \mathbf{s}_k(a)$ for $a \geq 1$, represents the $a$-th AP information in $\mathcal{S}_k$. The embedding vector is defined as the hidden state vector of the last block in the last layer, i.e., $\mathbf{v}_k = \mathbf{h}_{|\mathcal{A}_k|}^{(\xi)} |_{\mathrm{SENN}(\mathcal{S}_k|\theta^{\mu}_{\mathrm{SE}})}$.
Note that in the SENN, we use FNNs to embed the input and output. The notations of these FNNs are simplified from the above LSTM expressions, and their structures will be explained in Section \ref{sec:simulation_results}.

\subsubsection{Contention and Interference Predictors}
Using STA embedding vectors, we can predict how likely each STA pair is contending or hidden using NNs, which assists in deciding edge values. For instance, we use two fully connected NNs, the PCNN and the PHNN, to predict the indicators of contending and hidden pairs, $O^{\mathrm{C}}_{i,j}$ and $O^{\mathrm{H}}_{i,j}$, based on the embedding vectors $\mathbf{v}_i$ and $\mathbf{v}_j$ as
\begin{equation}\label{eq:predict_contention_and_interference}
\begin{aligned}
\tilde{O}^{\mathrm{C}}_{i,j}  = \mathrm{PCNN}(\mathbf{v}_i,\mathbf{v}_j|\theta^{\mu}_{\mathrm{PC}}) ;\
\tilde{O}^{\mathrm{H}}_{i,j}  = \mathrm{PHNN}(\mathbf{v}_i,\mathbf{v}_j|\theta^{\mu}_{\mathrm{PH}}) ,
\end{aligned}
\end{equation}
where $\tilde{O}^{\mathrm{C}}_{i,j}$ and $\tilde{O}^{\mathrm{H}}_{i,j}$ are the predicted values of $O^{\mathrm{C}}_{i,j}$ and $O^{\mathrm{H}}_{i,j}$, respectively, $\forall i \neq j$. 

\subsubsection{Edge Generator}
We extract key information on contention and interference to assist the IGL NN in deciding edge values. Specifically, for a given STA pair $i$ and $j$ (where $i \neq j$), the edge value $E_{i,j}$ is generated based on the information in $\mathcal{S}_i$ and $\mathcal{S}_j$, representing the negative impact of STA $i$'s transmissions on those of STA $j$ \cite{gu2024graph}, as discussed in Section \ref{subsec:igl_framework}.
We design the edge generator NN (EGNN) based on the extracted contention and interference information as
\begin{equation}\label{eq:eg_nn_structure}
\begin{aligned}
E_{i,j} = \mathrm{EGNN}(s_{i,\hat{a}_i},s_{i,\hat{a}_j},s_{j,\hat{a}_j},\tilde{O}^{\mathrm{C}}_{i,j},\tilde{O}^{\mathrm{H}}_{i,j}|\theta^\mu_{\mathrm{EG}}), \ \forall i\neq j  ,
\end{aligned}
\end{equation}
where $\tilde{O}^{\mathrm{C}}_{i,j}$ and $\tilde{O}^{\mathrm{H}}_{i,j}$ are predicted based on their state embeddings as \eqref{eq:predict_contention_and_interference}.
Note that the EGNN is configured to output a value in $[0,1]$, where the edge value is rounded to the nearest integer, $0$ or $1$, without explicit notation for simplicity.

In summary, the IGL NN parameters consist of the SENN parameters $\theta^{\mu}_{\mathrm{SE}}$, the predictor parameters $\theta^{\mu}_{\mathrm{PC}}$ and $\theta^{\mu}_{\mathrm{PH}}$, and the edge generator parameters $\theta^{\mu}_{\mathrm{EG}}$. When generating the edge for an STA pair $(i,j)$, $\forall i \neq j$, the IGL NN first embeds the STA states $\mathcal{S}_i$ and $\mathcal{S}_j$ using the SENN as in \eqref{eq:tokenizing_state}. It then predicts the contention and interference indicators based on the embeddings using the PCNN and the PHNN as in \eqref{eq:predict_contention_and_interference}, and finally generates the edge using the EGNN as in \eqref{eq:eg_nn_structure}.

\subsection{Evolution Strategy for Training the IGL NN}\label{subsec:es_egnn}

\subsubsection{Overview of ES}
The ES iterates in the following steps: 1) generate a new candidate IGL NN by adding Gaussian noise to the current parameters; 2) receive a reward signal given this candidate NN; 2) approximate the update direction of the current parameters and the noise variances according to the reward signal.
Since the ES algorithm is a parameter space algorithm, we pre-train some parts of the IGL NN to reduce the number of parameters of the NN to be trained by the ES. Specifically, the state embedding and the predictors are pre-trained using unsupervised learning and supervised learning, respectively, as explained in Appendix C. Consequently, these parameters in the IGL NN are fixed after pre-training, while only the parameters in the EGNN are updated by the ES.
The detailed implementation of the ES is presented as follows.

\begin{figure}[!t]
\centering
\includegraphics[scale=0.9]{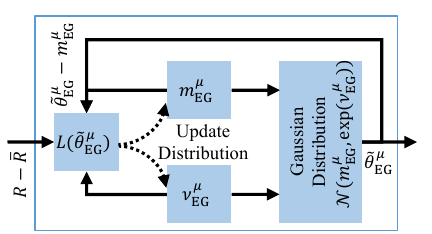}
\vspace{-0.2cm}
\caption{The ES algorithm that trains the EGNN in the IGL NN.}
\label{fig:es_flow}
\vspace{-0.25cm}
\end{figure}

\subsubsection{Network-Wise Reward Design}
We aggregate the STAs' reliability and the number of slots into one reward signal. Specifically, we augment the reliability definition as the number of packets transmitted within the minimum number of slots. 
For example, if the number of slots $Z$ obtained by coloring the graph exceeds the minimum $Z^*$, we can emulate an imaginary system that randomly selects STAs assigned to the $Z^*$ slots out of the total $Z$ slots to transmit, while STAs' transmissions in other slots are discarded. 
Consequently, the augmented reliability is expressed in the log scale as
\begin{equation}\label{eq:reward_signal}
\begin{aligned}
R = \begin{cases}
    \log(  \frac{Z^*}{Z} ) , \ \text{if } r_k \geq \hat{r},\ \forall k , \\
    \log( \min\{\frac{Z^*}{Z} , 1\}\cdot \frac{1}{K} \sum_{k=1}^{K} \min\{\frac{r_k}{\hat{r}} , 1\} ) , \ 
    \text{otherwise} .
\end{cases}
\end{aligned}
\end{equation}
If the reliability constraints are satisfied in the first case, the reward focuses on minimizing the number of slots; otherwise, when the number of slots used is less than the optimal number of slots, e.g., the NN outputs less edge than the optimal graph, the reward emphasizes satisfying the reliability constraints by guiding the algorithm to add more edges and separate more contending or interfering pairs.
Note that in the actual system, we do not need to implement the imaginary system that discards the STAs, and it is only for analysis to derive the reward signal.
Ideally, we can construct this reward signal if we know $Z^*$. Then, the maximum reward signal in \eqref{eq:reward_signal} is 0, and it is achieved if and only if the slot assignments minimize the number of slots while ensuring STA reliability.
However, note that the minimum number of slots $Z^*$ is unknown. Thus, when computing $R$, we use the chromatic number of the contention and interference graph, with edges defined as $(O^{\mathrm{C}}_{i,j} + O^{\mathrm{H}}_{i,j})$ for $i \neq j$, to approximate $Z^*$ as
\begin{equation}\label{eq:min_z_approximation}
\begin{aligned}
Z^*\!\approx\!\tilde{Z}^*\!=\!\chi([E_{i,j}|E_{i,j}=O^{\mathrm{C}}_{i,j} + O^{\mathrm{H}}_{i,j},\forall i\neq j;E_{k,k}=0,\forall k]),
\end{aligned}
\end{equation}
where $\chi(\cdot)$ returns the number of colors of the input graph's adjacency matrix using greedy coloring \cite{west2001introduction}. Since $Z^*\leq\tilde{Z}^*$, the approximated reward signal can be greater than $0$ and has an unknown maximum at $\log\frac{\tilde{Z}^*}{Z^*}$ (note that $\log\frac{\tilde{Z}^*}{Z^*}\geq0$). For simplicity, we use the same notation $R$ as the approximated reward signal in this paper.

To normalize the reward signal, we measure the average reward over the iterations as $\Bar{R}$ and aim to maximize the advantage \cite{west2001introduction,mnih2016asynchronous} of the reward with respect to the past average, as $R - \Bar{R}$.
Therefore, the loss of the EGNN parameters in the IGL is designed as
\begin{equation}
\begin{aligned}
L(\theta^\mu_{\mathrm{EG}}) = -\expt[R-\Bar{R}|\theta^\mu_{\mathrm{EG}}] .
\end{aligned}
\end{equation}

\subsubsection{Parameter Space Update Direction Estimation}
We optimize the EGNN parameters using a Gaussian ES algorithm \cite{salimans2017evolution}. In the ES algorithm, each parameter of the EGNN is assumed to be a Gaussian random variable, independent of each other, with the mean $m^\mu_{\mathrm{EG}}$ and the variance $\exp({\nu^\mu_{\mathrm{EG}}})$. Here, the exponential ensures the variance is always positive.
The ES optimizes the Gaussian-smoothed surrogate of the loss function as \cite{salimans2017evolution}
\begin{equation}\label{eq:es_loss_approximation}
\begin{aligned}
L(\theta^\mu_{\mathrm{EG}}) \approx \expt_{\tilde{\theta}^\mu_{\mathrm{EG}}\sim\mathcal{N}(m^\mu_{\mathrm{EG}},\exp({\nu^\mu_{\mathrm{EG}}}))}\big[L(\tilde{\theta}^\mu_{\mathrm{EG}})\big].
\end{aligned}
\end{equation}
The ES algorithm adjusts each parameter's mean and variance to find the optimal values that maximize the reward signal, as shown in Fig. \ref{fig:es_flow}. Specifically, it first randomly samples parameters $\tilde{\theta}^\mu_{\mathrm{EG}}$ from the current distribution as
\begin{equation}\label{eq:egnn_sample}
\begin{aligned}
\tilde{\theta}^\mu_{\mathrm{EG}} \sim  \mathcal{N}(m^\mu_{\mathrm{EG}},\exp({\nu^\mu_{\mathrm{EG}}})).
\end{aligned}
\end{equation}
Using the sampled EGNN parameters $\tilde{\theta}^\mu_{\mathrm{EG}}$, the IGL NN generates the edges in the graph based on the network states $\hat{\mathcal{S}}$, as described in Section \ref{subsec:ggm_structure}. The instant reward, $R$, is then evaluated as \eqref{eq:reward_signal} from the network based on the corresponding slot assignments by coloring the graph.
The gradient of the loss function with respect to the parameterized distribution (the mean and the variance) of the EGNN parameters based on \eqref{eq:es_loss_approximation} is
\begin{equation}
\begin{aligned}
&\ \ \ -\nabla_{m^\mu_{\mathrm{EG}},\nu^\mu_{\mathrm{EG}}} L(\theta^\mu_{\mathrm{EG}}) \\
&\approx -\nabla_{m^\mu_{\mathrm{EG}},\nu^\mu_{\mathrm{EG}}}  \expt_{\tilde{\theta}^\mu_{\mathrm{EG}}\sim\mathcal{N}(m^\mu_{\mathrm{EG}},\exp({\nu^\mu_{\mathrm{EG}}}))}\big[L(\tilde{\theta}^\mu_{\mathrm{EG}})\big] \\
&=\nabla_{m^\mu_{\mathrm{EG}},\nu^\mu_{\mathrm{EG}}}  \expt_{\tilde{\theta}^\mu_{\mathrm{EG}}\sim\mathcal{N}(m^\mu_{\mathrm{EG}},\exp({\nu^\mu_{\mathrm{EG}}}))}\big[\expt[R-\Bar{R}|\theta^\mu_{\mathrm{EG}}]\big] \\
&\stackrel{\text{(a)}}{=} \expt_{\tilde{\theta}^\mu_{\mathrm{EG}}\sim\mathcal{N}(m^\mu_{\mathrm{EG}},\exp({\nu^\mu_{\mathrm{EG}}}))}\big[\expt[R-\Bar{R}|\theta^\mu_{\mathrm{EG}}] \cdot \\
&\qquad\qquad\qquad\qquad\qquad\qquad\qquad  \nabla_{m^\mu_{\mathrm{EG}},\nu^\mu_{\mathrm{EG}}} \log p(\tilde{\theta}^\mu_{\mathrm{EG}})\big] \\
&\stackrel{\text{(b)}}{\approx}  (R-\Bar{R})\nabla_{m^\mu_{\mathrm{EG}},\nu^\mu_{\mathrm{EG}}}\log p(\tilde{\theta}^\mu_{\mathrm{EG}}), 
\end{aligned}
\end{equation}
where $p(\cdot)$ is the probability density function of the EGNN parameters, i.e., the Gaussian distribution with mean $m^\mu_{\mathrm{EG}}$ and variance $\exp({\nu^\mu_{\mathrm{EG}}})$.
Here, (a) uses the log-derivative trick to move the gradient computation into the expectation \cite{sutton1999policy}, and (b) is the approximation of the expected gradient in the current iteration.
By further substituting the expression for the Gaussian distribution's probability density function, we have
\begin{equation}\label{eq:egnn_update}
\begin{aligned}
-\nabla_{m^\mu_{\mathrm{EG}}} L(\theta^\mu_{\mathrm{EG}}) &\approx (R-\Bar{R}) \cdot \frac{\tilde{\theta}^\mu_{\mathrm{EG}}- m^\mu_{\mathrm{EG}}}{\exp({\nu^\mu_{\mathrm{EG}}})} ; \\
-\nabla_{\nu^\mu_{\mathrm{EG}}} L(\theta^\mu_{\mathrm{EG}}) &\approx (R-\Bar{R})\cdot ( \frac{(\tilde{\theta}^\mu_{\mathrm{EG}}- m^\mu_{\mathrm{EG}})^2}{2\exp({\nu^\mu_{\mathrm{EG}}})} - \frac{1}{2} ) .
\end{aligned}
\end{equation}
The complete algorithm using the ES will be summarized in the next section, along with the application of the DHF.

\section{Accelerated Training and Inference of IGL NN Using DHF in IGL}\label{sec:dhf}

\begin{figure}[!t]
\centering
\includegraphics[scale=0.9]{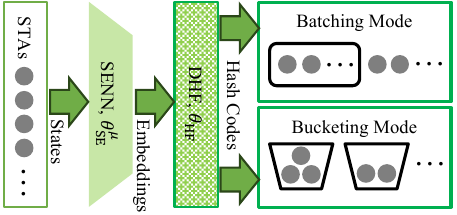}
\vspace{-0.2cm}
\caption{The batching mode and the bucketing model of the DHF, which selects a subset of contending and interfering STAs for the training and the inference of the IGL NN, respectively.}
\label{fig:hashing}
\vspace{-0.25cm}
\end{figure}

This section explains the DHF in the IGL that accelerates the IGL by focusing the training and inference of the IGL NN only on those contending and hidden STA pairs. We first explain the design of the DHF in Section \ref{subsec:dhf} and then show the usage of the DHF in the training and the inference of the IGL NN in Sections \ref{subsec:dhf_batching} and \ref{subsec:dhf_bucketing}, respectively.

\subsection{Design of Deep Hashing Function}\label{subsec:dhf}
We use a DHF NN to approximate the LSH, where the NN encodes each STA's state as a code vector. The similarity between the codes of two STAs approximates the geometric proximity and how likely the contention and interference will occur between them.
Since the exact location of STAs is not measured, we use the state embeddings of STA pairs as the input of hashing to approximate the geometric relation in each STA pair.
Then, by comparing the bits of the STAs' hash codes, we can efficiently find the contending and interfering STA pairs.
Specifically, the DHF NN with parameters $\theta_{\mathrm{HF}}$ to compute each STA's soft hash code based on the STA's state embeddings as
\begin{equation}\label{eq:hashing_token}
\begin{aligned}
\tilde{\mathbf{b}}_k\triangleq [\tilde{b}_k^1,\dots,\tilde{b}_k^\Lambda]  =  \mathrm{DHF}(\mathbf{v}_k|\theta_{\mathrm{HF}}), \ \forall k , 
\end{aligned}
\end{equation}
where $\tilde{b}_k^n \in [-1,1]$ for $n = 1, \dots, \Lambda$, and $\Lambda$ is the number of bits in each hash code.
To train the DHF, part of the loss is designed to adjust the similarity between codes based on the contention and interference indicators $O^{\mathrm{C}}_{i,j} + O^{\mathrm{H}}_{i,j}$ as
\begin{equation}\label{eq:loss_dhf_similarity}
\begin{aligned}
L'(\theta_{\mathrm{HF}}) = \frac{1}{|\mathcal{E}|}\sum_{i\neq j}\big[\frac{1}{2\Lambda}(\sum_{n=1}^{\Lambda} \tilde{b}_i^n\tilde{b}_j^n + \Lambda) - (O^{\mathrm{C}}_{i,j}+O^{\mathrm{H}}_{i,j})\big]^2 ,
\end{aligned}
\end{equation}
where $\frac{1}{2\Lambda} ( \sum_{n=1}^{\Lambda} \tilde{b}_i^n \tilde{b}_j^n + \Lambda )$ is the normalized Hamming distance between STAs $i$ and $j$'s hash codes. The loss in \eqref{eq:loss_dhf_similarity} reduces the Hamming distances between STAs' codes if they are contending or hidden from each other, or otherwise increases the Hamming distances.
Further, to ensure that each bit encodes different information about the STAs' states, part of the loss is designed to reduce the correlation between bit positions in the hash codes as \cite{luo2023survey}
\begin{equation}\label{eq:loss_dhf_cor}
\begin{aligned}
L''(\theta_{\mathrm{HF}}) = \frac{1}{K^2}\|[\tilde{\mathbf{b}}_1,\dots,\tilde{\mathbf{b}}_K][\tilde{\mathbf{b}}_1,\dots,\tilde{\mathbf{b}}_K]^{\rm T}-\mathbb{I}^{K\times K}\|_2^2,
\end{aligned}
\end{equation}
where $[\tilde{\mathbf{b}}_1, \dots, \tilde{\mathbf{b}}_K][\tilde{\mathbf{b}}_1, \dots, \tilde{\mathbf{b}}_K]^{\rm T}$ is the matrix with off-diagonal elements that measure the correlations between bit positions. The total loss of the DHF is
\begin{equation}\label{eq:loss_dhf_total}
\begin{aligned}
L(\theta_{\mathrm{HF}}) = L'(\theta_{\mathrm{HF}}) + \lambda L''(\theta_{\mathrm{HF}}) .
\end{aligned}
\end{equation}
where $\lambda > 0$ is the weight of the correlation loss part in the total loss. The training of the DHF is outlined in Algorithm \ref{alg:dhf}. 
The hard hash code is generated using the trained DHF as
\begin{equation}
\begin{aligned}
\mathbf{b}_k\triangleq [b_k^1,\dots,b_k^\Lambda] =  [\sgn \tilde{b}_k^1,\dots,\sgn \tilde{b}_k^\Lambda] ,\ \forall k,
\end{aligned}
\end{equation}
where $\sgn$ takes the sign, $-1$ or $1$, of the soft hard code bits.
The applications of the trained DHF are explained below.

\begin{algorithm}[!t]
\caption{Training of the DHF}\label{alg:dhf}
\begin{algorithmic}[1]
\STATE Set training steps $M$ and the learning rate $\eta$.\label{alg:line:dhf:start}
\STATE Randomly initialize the DHF parameters, $\theta_{\mathrm{HF}}$.
\FOR{step $1,\dots,M$}
\STATE Randomly generate a network with $K$ STAs.
\STATE Embed STA states using the trained SENN as \eqref{eq:tokenizing_state}.
\STATE Compute the hash codes of STA embeddings as \eqref{eq:hashing_token}.
\STATE Compute the losses $L(\theta_{\mathrm{HF}})$ as \eqref{eq:loss_dhf_similarity}\eqref{eq:loss_dhf_cor}\eqref{eq:loss_dhf_total}.
\STATE Update the parameters of the DHF as \\
\quad\quad\quad $\theta_{\mathrm{HF}}= \theta_{\mathrm{HF}}-\eta\nabla_{\theta_{\mathrm{HF}}}L(\theta_{\mathrm{HF}})$.
\ENDFOR
\STATE \textbf{return} the trained DHF with $\theta_{\mathrm{HF}}$.\label{alg:line:dhf:end}
\end{algorithmic}
\end{algorithm}

\subsection{STA Batching Mode of DHF}\label{subsec:dhf_batching}
Based on the hash codes generated by the DHF, we can select a small subset of STAs with a high likelihood of contention and interference for the ES algorithm, reducing the training time.

\subsubsection{Design of Batching Mode}
The batching mode starts by randomly choosing $\Psi$ positions in $n = 1, \dots, \Lambda$ without replacement as $n'_1, \dots, n'_\Psi$. Then, a $\Psi$-bit random binary query code is generated as $q_1, \dots, q_\Psi$. STAs are selected if their hash codes at $n'_1, \dots, n'_\Psi$ match the random query codes as
\begin{equation}
\begin{aligned}
\mathcal{Q} \leftarrow \mathcal{Q} \cup \{k\in \mathcal{V}| [b_k^{n'_1},\dots,b_k^{n'_\Psi}] = [q_1,\dots,q_\Psi]\} ,
\end{aligned}
\end{equation}
where $\mathcal{Q}$ represents the selected STAs and is initialized as $\emptyset$. The above process is repeated with a new query code at new random positions in each iteration until $\mathcal{Q}$ contains $K'$ STAs. 
We refer to $\mathcal{Q}$ as the batch of STAs and $K' = |\mathcal{Q}|$ as the batch size.
Since the DHF is trained to output the same bits in STA hash codes if STAs are contending or hidden, as explained in Section \ref{subsec:dhf}, STAs that are likely contending with and hidden from each other will likely match the query codes and be batched together. This STA batching mode is used when the IGL NN is under training to select a subset of STAs, allowing the training algorithm to focus on those contending and hidden pairs, thereby increasing efficiency.

\subsubsection{ES with Batching Mode of DHF}
To reduce the computing time of the training process of the ES, we design a curriculum learning method \cite{wang2022survey} using the batching mode of the DHF to gradually increase the number of STAs, i.e., raise the difficulty of the graph-constructing task, until it reaches the total number of STAs. Specifically, as shown in Algorithm \ref{alg:es-ggm}, the ES algorithm starts with a small number of STAs, $K'$, and the total number of STAs, $K$.
Here, the state embedding, the predictors, and the DHF are pre-trained. The distribution of the edge generator parameters is initialized with the mean $m^\mu_{\mathrm{EG}} = 0$ and the variance $\exp({\nu^\mu_{\mathrm{EG}}}) = \sigma^2_\mathrm{0}$.
In each ES iteration, a random Wi-Fi network is simulated, and $K'$ STAs are selected using the DHF's batching mode. After that, the EGNN parameters are sampled from the parameterized distribution, as shown in \eqref{eq:egnn_sample}, which constructs the IGL NN with the pre-trained state embedding and predictors, as shown in Fig. \ref{fig:ggm_structure} of Section \ref{subsec:ggm_structure}.
For the given STA batch, the edges between them are generated using the IGL NN as \eqref{eq:tokenizing_state}, \eqref{eq:predict_contention_and_interference}, and \eqref{eq:eg_nn_structure}, forming the graph. Next, the graph is colored, and the slot assignments for the batched STAs are determined accordingly. The assignments are applied in the network, with only the batched STAs transmitting, where the transmission reliabilities are measured, and the reward is computed as in \eqref{eq:reward_signal}.
Finally, the EGNN parameters' distribution is updated in the direction indicated in \eqref{eq:egnn_update}, with $\eta$ as the learning rate. At the end of each iteration, the moving average of the previous reward performance is measured as
\begin{equation}\label{eq:averaged_reward_performance}
\begin{aligned}
\Omega \leftarrow \gamma \Omega + (1-\gamma) \mathbf{1}_{\{R\geq0\}} ,
\end{aligned}
\end{equation}
where $\gamma$ is a smoothing factor less than 1. Since the graph's performance is near-optimal when $R \geq 0$ (note that the maximum $R$ is greater than $0$ as an approximation of the minimum slots is used in \eqref{eq:reward_signal}\eqref{eq:min_z_approximation}), we use $\mathbf{1}_{\{R \geq 0\}}$ as an indicator of whether the IGL NN has satisfactory performance.
In \eqref{eq:averaged_reward_performance}, if the averaged reward indicator $\Omega$ is greater than a threshold $\hat{\Omega}$, i.e., the IGL NN achieves an averaged satisfactory performance with the current number of STAs $K'$, then the number of batched STAs $K'$ increases by $K''$ in the next iteration; otherwise, the algorithm keeps the same number of STAs $K'$. The algorithm terminates if it achieves an averaged satisfactory performance with $K$ STAs. The mean of the parameter distribution is returned as the trained EGNN parameters.

\begin{algorithm}[!t]
\caption{ES With Batching Mode of the DHF}\label{alg:es-ggm}
\begin{algorithmic}[1]
\STATE Set the batch size $K'$ and the total STA number $K$.
\STATE Set the learning rate $\eta$ and batch increment step $K''$.
\STATE Train $\theta^{\mu}_{\mathrm{SE}}$, $\theta^{\mu}_{\mathrm{PC}}$ and $\theta^{\mu}_{\mathrm{PH}}$ as in Appendix C.
\STATE Train $\theta_{\mathrm{HF}}$ as Algorithm \ref{alg:dhf}.
\STATE Initialize $m^\mu_{\mathrm{EG}}=0$ and $\exp({\nu^\mu_{\mathrm{EG}}})=  \sigma^2_\mathrm{0}$.
\FOR{step $1,2,\dots$}
    \STATE Set up a Wi-Fi network with $K$ STAs.
    \STATE Select $K'$ STAs using the batching mode of the DHF.
    \STATE Sample the EGNN parameters $\tilde{\theta}^{\mu}_{\mathrm{EG}}$ as \eqref{eq:egnn_sample}.
    \STATE Generate edges in the batch as \eqref{eq:tokenizing_state}, \eqref{eq:predict_contention_and_interference} and \eqref{eq:eg_nn_structure}.
    \STATE Color the graph using greedy coloring, $\mathbf{z},Z=\chi(\mathbf{E})$.
    \STATE Apply the slot assignments as $\mathbf{z}$ and $Z$.
    \STATE Measure reliabilities and obtain the reward $R$ as \eqref{eq:reward_signal}.
    \STATE Estimate update directions of $m^\mu_{\mathrm{EG}}$ and $\nu^\mu_{\mathrm{EG}}$ in \eqref{eq:egnn_update}.
    \STATE Update $m^\mu_{\mathrm{EG}}$ and $\nu^\mu_{\mathrm{EG}}$ as \\
    \quad $m^\mu_{\mathrm{EG}}\leftarrow m^\mu_{\mathrm{EG}} - \eta\nabla_{m^\mu_{\mathrm{EG}}} L(\theta^\mu_{\mathrm{EG}})$; \\
    \quad\ $\nu^\mu_{\mathrm{EG}}\leftarrow \nu^\mu_{\mathrm{EG}} - \eta\nabla_{\nu^\mu_{\mathrm{EG}}} L(\theta^\mu_{\mathrm{EG}})$.
    \STATE Evaluate the reward performance indicator $\Omega$ as \eqref{eq:averaged_reward_performance}.
    \IF{$\Omega\geq\hat{\Omega}$}
        \STATE \textbf{if} $K'' = K$ \textbf{then} \textbf{break}.
        \STATE \textbf{else} $K' \leftarrow \min\{K' + K'',K\}$.
    \ENDIF
\ENDFOR
\STATE \textbf{return} the trained EGNN with $\theta^{\mu}_{\mathrm{EG}}=m^\mu_{\mathrm{EG}}$.
\end{algorithmic}
\end{algorithm}

\subsection{STA Bucketing Mode of DHF}\label{subsec:dhf_bucketing}
Furthermore, we use the DHF to bucket those highly likely contending and hidden STA pairs and generate edges only between them to reduce the computing time of the inference of the trained IGL NN.
\subsubsection{Design of Bucketing Mode}
The bucketing mode randomly chooses $\Psi$ positions in $n = 1, \dots, \Lambda$ without replacement as $n'_1, \dots, n'_\Psi$. A hash table is constructed based on STAs' hash codes at the chosen positions. Let $\mathcal{C}$ be the set of all hash codes at the selected positions in the table as
\begin{equation}\label{eq:bucketing_code_generation}
\begin{aligned}
\mathcal{C} \leftarrow \cup_k \{[b_k^{n'_1},\dots,b_k^{n'_\Psi}]\} . 
\end{aligned}
\end{equation}
Then, for every code in $\mathcal{C}$, a bucket of STAs is collected if they have the same codes in the selected positions as
\begin{equation}\label{eq:bucketing}
\begin{aligned}
\mathcal{B}(\mathbf{q}) \leftarrow \{k\in\mathcal{V}| [b_k^{n'_1},\dots,b_k^{n'_\Psi}] = \mathbf{q} \}, \ \forall \mathbf{q}\in \mathcal{C} ,
\end{aligned}
\end{equation}
The above process in \eqref{eq:bucketing_code_generation} and \eqref{eq:bucketing} is repeated to construct $\Upsilon$ hash tables, bucketing a sufficient number of STA pairs for edge generation as
\begin{equation}\label{eq:repeated_bucketing}
\begin{aligned}
\mathcal{P} \leftarrow \mathcal{P} \cup \left(\cup_{\mathbf{q}\in \mathcal{C}} \{ (i,j) | i \in \mathcal{B}(\mathbf{q})\land j \in \mathcal{B}(\mathbf{q}) , i\neq j \}\right),
\end{aligned}
\end{equation}
where new random $\Psi$ positions are chosen in each repetition. Here, $\mathcal{P}$ represents the STA pairs requiring IGL NN operations and is initialized as $\emptyset$. The edges between STAs in the same bucket are generated using the IGL NN, as described in Section \ref{subsec:ggm_structure}, while the edge values for those non-bucketed pairs are ignored and are set to $0$, as
\begin{equation}\label{eq:eg_in_bucket}
\begin{aligned}
E_{i,j} \leftarrow \mu(\mathcal{S}_i,\mathcal{S}_j|\theta^\mu),\forall(i,j) \in \mathcal{P};\
E_{i,j} \leftarrow 0,\forall (i,j)\notin \mathcal{P} .
\end{aligned}
\end{equation}
Since the hash codes of contending and hidden STAs are close, they are more likely to be grouped in the same bucket, and edges between them are processed, while edges between less likely contending or hidden STAs are ignored.
This STA bucketing mode is used when the trained IGL NN is deployed online across the entire network to reduce the processed edges, i.e., to reduce the complexity of the inference of the IGL NN.

\subsubsection{Online Graph Construction with Bucketing Model of DHF}
\begin{figure}[!t]
\centering
\includegraphics[scale=0.9]{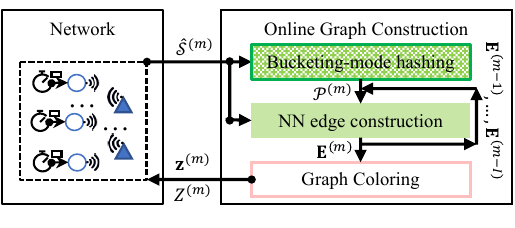}
\vspace{-0.2cm}
\caption{Illustration of the online graph construction architecture.}
\label{fig:online_architecture}
\vspace{-0.25cm}
\end{figure}
We then present the online graph construction architecture that deploys the IGL NN with the bucketing model of the DHF, as shown in Fig. \ref{fig:online_architecture}. 
Note that we assume the IGL NN has been trained and its parameter is not updated further. 
In this architecture, the online graph construction process outputs the graph and slot assignments according to online-measured network states and runs in parallel with the network.

In the $m$-th round of the online process, $m=1,2,\dots$, the network measures the states $\hat{\mathcal{S}}^{(m)}$, e.g., based on the previous transmissions. Then, the process uses the bucketing mode of the DHF to identify the STA pairs likely contending or interfering, indicated in $\mathcal{P}^{(m)}$, as shown in \eqref{eq:repeated_bucketing}. Since the bucketing mode may miss some STA pairs, the process combines $\mathcal{P}^{(m)}$ with STA pairs that had edges generated in the previous $I$ rounds as
\begin{equation}\label{eq:online_sta_pair_augment}
\begin{aligned}
\mathcal{P}^{(m)} \leftarrow \mathcal{P}^{(m)} \cup \big(\cup_{m'=1,\dots,I}\{(i,j)|E_{i,j}^{(m-m')}=1\}\big) .
\end{aligned}
\end{equation}
This is because previously generated edges are more likely to be generated in the following rounds. Here, $E_{i,j}^{(m-m')}$ in \eqref{eq:online_sta_pair_augment} represents the edge value between STAs $i$ and $j$ in the $(m-m')$-th round.
The IGL then generates the edges between each of these pairs, forming the graph with the adjacency matrix $\mathbf{E}^{(m)}$. The slot assignments, $\mathbf{z}^{(m)}$ and $Z^{(m)}$, are determined by coloring this graph. The process takes a duration of $\tau^{(m)}$ milliseconds for the process to perform the bucketing hashing, the graph construction, and the slot assignments in total.
Note that the network makes no transmissions during the first round of the online process, i.e., during the first $\tau^{(1)}$ milliseconds. After the slot assignments are returned from the IGL process to the network, the network starts STA transmissions with the assignments, and the process begins the next round of assignments. The above process repeats until the network stops.

\section{Simulation Results}\label{sec:simulation_results}
This section evaluates our proposed methods in simulations.
\subsection{Simulation Configurations}\label{subsec:ns3_simulation_config}
We use NS-3 \cite{nsnam} to simulate Wi-Fi networks with the RTWT mechanism. 
All devices are located in a 100m$\times$100m squared area centered at the coordinates $(50,50)$ meters, simulating a factory environment \cite{3gpp.38.825}.
We assume the network has $A=100$ APs, and they are located at the grid in the simulated area, i.e., they are at the coordinates $(5+10x,5+10y)$ meters, where $(x,y)$ are the indices of APs in $x$ and $y$ axes, i.e., $x=1,\dots,10$ and $y=1,\dots,10$.
Unless specified otherwise, STAs are static, and the total number of STAs is $K = 1000$. Each STA is randomly distributed within the simulated area, with each coordinate selected randomly from a uniform distribution within the interval $[0, 100]$ meters. The duration of a RTWT slot is $\Delta_\mathrm{0} = 500$ microseconds. The channel bandwidth is $B = 20$ MHz at a 5.8-GHz carrier frequency.
The transmission power of STAs is $\mathbf{P}_\mathrm{0}=0$~dBm, and the noise power is set as $\mathbb{N}_\mathrm{0}B =-96$~dBm.
The path losses between any two devices (including all APs and STAs) follow a log-distance path loss model \cite{series2015propagation} as $-28\log_{10}(l+1)-20\log_{10}(f/(1\text{MHz}))+12$ in decibel scale, where $f$ is the carrier frequency in MHz and $l$ is the distance between two devices in meters.
The receiver sensitivity threshold is set (in decibel scale) to $\mathbf{P}_\mathrm{0}-s_{\max}=-95$~dBm. 
The packet size $L$ is 800 bits.
The actual decoding error probability of each transmission is computed using the same equation in \eqref{eq:tx_error_for_mcs_wi-fi}, where $\phi$ is the signal-to-interference-plus-noise of each transmission instead. This is implemented as a transmission error model in NS-3. The target reliability of each STA is $\hat{r}=0.99$. The maximum number of retransmissions for each STA in a RTWT slot is $5$.

The dimensions of the embedding vectors $\mathbf{v}_k$ $\forall k$ and the hash codes $\mathbf{b}_k$ $\forall k$ are $5$ and $30$, respectively. The configurations of all FNNs are listed in Table \ref{tab:configurations_of_FNNs}, including the size of each layer, the hidden layer activation functions (HAFs), and the output activation functions (OAFs). Here, the FNNs at the input and output of the SENN are denoted as SENNI and SENNO, respectively.
The number of LSTM layers in the state embedding and the decoder is $\xi = 2$. The weight of the correlation loss in the DHF's total loss is $\lambda = 0.2$. 
The learning rate is $\eta = 10^{-3}$ for training the state embedding, the predictors, and the DHF, and the corresponding training steps $M$ are $2000$, $2000$, and $10000$, respectively. 
For training of the EGNN using the ES, the smoothing factor in \eqref{eq:averaged_reward_performance} is $\gamma = 0.9$; the threshold for the reward performance to increase STAs is $\hat{\Omega} = 0.9$; the number of STAs added in each increment is $K'' = 50$; the learning rate is $\eta = 10^{-1}$. The initial variance of the EGNN parameters is $\sigma^2_\mathrm{0} = 0.1$.
Simulations are conducted on a MacBook with the M4 chip.

\begin{table}[t]
\caption{Configurations of All FNNs}
\label{tab:configurations_of_FNNs}
\begin{minipage}{\columnwidth}
\begin{center}
\begin{tabular}{|c|c|c|c|}
\hline
NNs     & Dimensions of layers      & HAFs                      & OAFs \\ \hline
SENNI   &   $3,15,15,15$            & $\mathrm{GELU}(\cdot)$    & $\mathrm{GELU}(\cdot)$               \\ \hline
SENNO   &   $15,5$                  & None                      & None \\ \hline
PCNN    &   $10,50,50,1$            & $\mathrm{ReLU}(\cdot)$    & $\mathrm{Sigmoid}(\cdot)$               \\ \hline
PHNN    &   $10,50,50,1$            & $\mathrm{ReLU}(\cdot)$    & $\mathrm{Sigmoid}(\cdot)$               \\ \hline
DHF     &   $5,30,30,30,30,30$      & $\mathrm{GELU}(\cdot)$    & $\mathrm{Tanh}(\cdot)$          \\ \hline
EGNN    &   $5,50,50,1$          & $\mathrm{ReLU}(\cdot)$ & $\mathrm{Sigmoid}(\cdot)$            \\ \hline
\end{tabular}
\vspace{-0.2cm}
\end{center}
\end{minipage}
\end{table}

\subsection{Baseline Methods}
We explain the baseline methods, other than the proposed methods. Note that since contention and interference are not measured from the network, the Markov models can hardly estimate the STA QoS. Additionally, the GNN models can hardly provide useful slot assignment decisions in graph-coloring-structured problems based on previous observations in \cite{gu2024graph} and justifications in \cite{xu2018how,loukas2019what}. Thus, we compare the remaining approaches reviewed in Section \ref{subsec:related_works}, including human intuition-based graph models and the IGL with PG or DPG. Their detailed implementations are explained as follows.

\subsubsection{Human Intuition-Based Graph Models (IFG, CHG)}
We compare the human intuition-based graph models used to construct binary graphs with manually designed rules. The work in \cite{chen2022energy} connects two STAs if the same AP can detect them, i.e., $E_{i,j} = \mathbf{1}_{\{\mathcal{A}_i \cap \mathcal{A}_j \neq \emptyset \}}$ for all $i \neq j$, referred to as the ``IFG'' scheme.
Since STAs' transmissions are primarily affected by contention and interference in the network, we also compare the graph constructed using the contention and interference indicators, where $E_{i,j} = O^{\mathrm{C}}_{i,j} + O^{\mathrm{H}}_{i,j}$ for all $i \neq j$, referred to as the ``CHG'' scheme.
Note that the CHG scheme is applied for comparison and is impractical in the real world due to the unmeasured contention and interference indicators.

\subsubsection{IGL Requiring Edge-Wise Feedback (PG, DPG)}
We compare the designed ES algorithm with the PG \cite{sutton1999policy} and DPG \cite{lillicrap2019continuous} algorithms when training the IGL NN.
Specifically, the PG algorithm uses the log-derivative trick to estimate the gradient of the performance w.r.t. each edge value in \eqref{eq:chain_rule_for_graph} as
\begin{equation}
\begin{aligned}
\frac{\partial L(\theta^\mu)}{\partial E_{i,j}} = \frac{(R-\Bar{R})\partial \log E_{i,j} }{\partial E_{i,j}} , \ \forall i\neq j , 
\end{aligned}
\end{equation}
where $E_{i,j} \in [0,1]$ is interpreted as the probability that an edge exists between STA $i$ and STA $j$. 
For the DPG, the gradient w.r.t. the edges are estimated based on the back-propagation on another NN that approximates the performance given edge values and STA states as
\begin{equation}
\begin{aligned}
\frac{\partial L(\theta^\mu)}{\partial E_{i,j}} = \frac{\partial Q(\mathbf{E},\hat{\mathcal{S}}|\theta^Q) }{\partial E_{i,j}}, \ \forall i\neq j , 
\end{aligned}
\end{equation}
where $Q(\cdot|\theta^Q)$, referred to as the critic, is trained to minimize the difference between its output and the reward \cite{gu2024graph}.

\subsection{Training Convergence of SENN, PCNN, PHNN and DHF}
\begin{figure}[!t]
\centering
\subfloat[Pre-training of the IGL NN components.]{\includegraphics[scale=0.9]{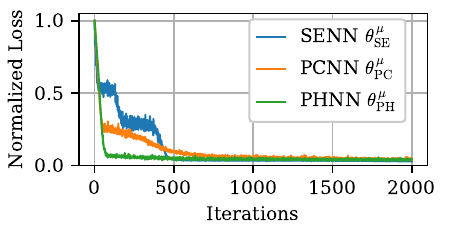}\label{fig:plot_pretraining_ggm}}\\
\vspace{-0.2cm}
\subfloat[Training of the DHF.]{\includegraphics[scale=0.9]{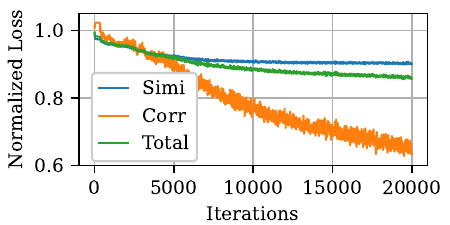}
\label{fig:plot_pretraining_dhf}}
\caption{Normalized loss values during training processes of the SENN, the PCNN, the PHNN and the DHF.}
\label{fig:plot_pretraining}
\vspace{-0.25cm}
\end{figure}
We first show the loss values in the training process of the SENN, the PCNN and the PHNN in Fig. \ref{fig:plot_pretraining_ggm}, where the loss values are normalized against their initial values during training. Results show that the loss values decrease over the training steps and converge in less than 500 steps when training the IGL NN components, the state embedding and the predictors.
Fig. \ref{fig:plot_pretraining_dhf} shows that the DHF's losses, i.e., the similarity loss (with legend ``Simi''), the correlation loss (with legend ``Corr'') and the total loss (with legend ``Total''). The similarity loss converges over the iteration, while the correlation and total loss constantly decreases. We use the NNs trained at $20000$ steps, including the state embedding, the predictors, and the DHF, in the following simulations without further updates to their parameters.
Note that we show the training losses in Fig. \ref{fig:plot_pretraining} to demonstrate the convergence of the training process, while the validation performance of the trained NNs is shown in the subsequent simulation results.

\subsection{Configurations of Batching/Bucketing Modes of DHF}\label{subsec:simulation_dhf}
\begin{figure}[!t]
\centering
\subfloat[The reward of the batching mode with different numbers of query bits, $\Psi$, when $K'=20$.]{\includegraphics[scale=0.9]{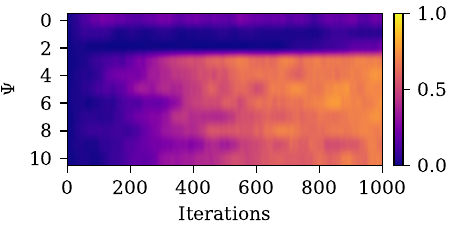}\label{fig:plot_rwd_es_with_dhf_nqbit}}\\
\vspace{-0.2cm}
\subfloat[The performance of the bucketing mode with different numbers of query bits, $\Psi$, and hash tables, $\Upsilon$.]{\includegraphics[scale=0.9]{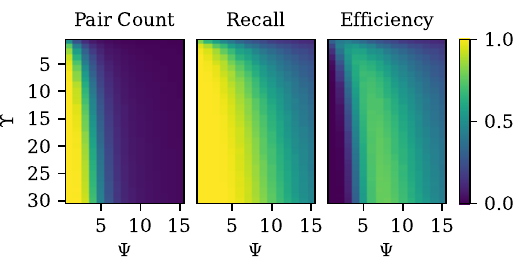}
\label{fig:plot_dhf_bit_tab_grid}}
\caption{The performance of the trained DHF.}
\label{fig:plot_dhf}
\vspace{-0.25cm}
\end{figure}

Based on the trained DHF, we then test the batching and bucketing modes. First, for the batching mode, we run the ES algorithm with a fixed small number of STAs, $K' = 20$, selected (i.e., without curriculum learning). The reward performance is measured as the moving average of the indicator $\mathbf{1}_{R \geq 0}$ at each step. Fig. \ref{fig:plot_rwd_es_with_dhf_nqbit} compares the performance of the batching mode when using different numbers of query bits, from $\Psi = 0$ to $10$, where $\Psi = 0$ means the STAs are chosen randomly without using the DHF.
The results indicate that the ES algorithm struggles to converge when using fewer query bits, e.g., $\Psi = 0, 1, 2$. This is because the selected STAs are independent of their contention and interference, and they are likely to have no contention or interference between them. As a result, the IGL NN finds it difficult to learn to separate contending or hidden STAs. The convergence improves when $\Psi = 3, 4$. This is because more query bits lead to a higher proportion of contending or interfering STA pairs in the selected batch, which enhances training efficiency. However, the convergence slows slightly when $\Psi > 4$ because the batch becomes highly biased toward contending or hidden STA pairs, and the IGL NN struggles to learn when not to separate STA pairs within the batch.
Second, for the bucketing mode, we measure the proportion of hashed STA pairs from all possible STA pairs (with legend ``Pair Count''), and the proportion of hashed contending or hidden pairs from all contending or hidden pairs (with legend ``Recall''). Fig. \ref{fig:plot_dhf_bit_tab_grid} compares the above metrics for different numbers of query bits, from $\Psi = 1$ to $15$, and hash tables, from $\Upsilon = 1$ to $30$.
The results show that with fewer query bits and more hash tables, more STA pairs are hashed, and the recall rate is higher. Since we aim to reduce the number of hashed STA pairs while increasing the number of hashed contending or hidden pairs among all hashed pairs, we subtract ``Pair Count'' from ``Recall'' to measure the efficiency of the bucketing mode (with legend ``Efficiency''). The results indicate that $\Psi \approx 7$ and $\Upsilon \geq 20$ achieve the best efficiency.
In summary, we fix $\Psi = 4$ in the batching mode and $\Psi=7$ and $\Upsilon=20$ in the bucketing mode in the rest of the simulations.

\subsection{Performance of ES with Batching Mode of DHF}
Next, we illustrate the ES algorithm's performance.
\subsubsection{Comparison with PG and DPG}
\begin{figure}[!t]
\centering
\includegraphics[scale=0.9]{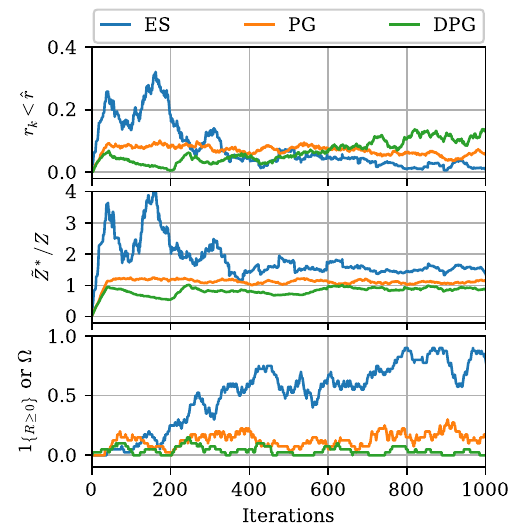}
\vspace{-0.2cm}
\caption{The performance of ES, PG and DPG for the IGL when $K'=20$.}
\label{fig:plot_rwd_nc_q_es_pg_dpg}
\vspace{-0.25cm}
\end{figure}

We compare the performance of the ES with the PG and the DPG when a fixed small number of STAs is selected using the batching mode of the DHF, i.e., $K' = 20$. The performance is measured in Fig. \ref{fig:plot_rwd_nc_q_es_pg_dpg} based on the QoS violations, the ratio of approximated optimal slots in \eqref{eq:min_z_approximation} to the number of used slots, and the averaged reward performance indicator. The results show that the proposed ES algorithm achieves the lowest QoS violation probability, the least number of slots, and the highest reward performance (approximately $4\sim10$ times higher), while the PG and DPG algorithms struggle to converge to satisfactory performance. This is because the PG and DPG algorithms require STA-pairwise (or edge-wise) gradients, but no STA-pairwise feedback is available. In comparison, the ES algorithm directly applies the reward feedback in the parameter space to improve the IGL NN.

\begin{figure}[!t]
\centering
\includegraphics[scale=0.9]{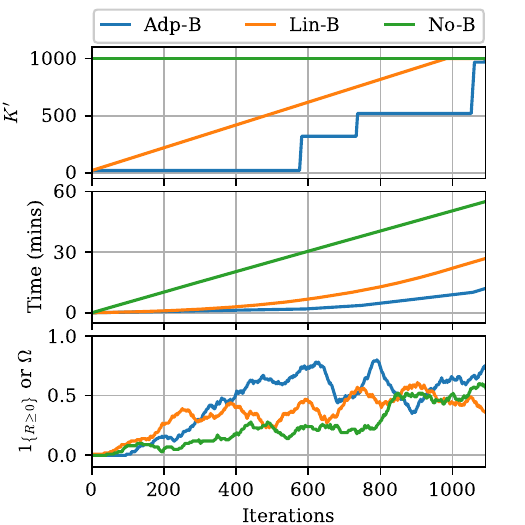}
\vspace{-0.2cm}
\caption{The performance of the ES with different batching schemes.}
\label{fig:plot_cl}
\vspace{-0.25cm}
\end{figure}

\subsubsection{Performance of ES with Different Batching Schemes}
We then show the performance of the ES algorithm with and without the batching mode of the DHF. Specifically, we consider 1) the proposed batching method that adaptively increases the number of STAs, $K'$, based on the reward performance (with legend ``Adp-B''), as shown in Algorithm \ref{alg:es-ggm}; 2) a batching method that linearly increases $K'$ by one in each step (with the legend ``Lin-B''); and 3) a method that uses no batching but trains the IGL NN with the total number of STAs, $K = K'$ (with the legend ``No-B''). Fig. \ref{fig:plot_cl} measures the number of STAs, the training time, and the averaged reward performance indicator during the training steps.
The results show that the proposed adaptive batching converges to the target number of STAs at approximately $1100$ steps. Additionally, the performance of all three compared methods is similar around $1100$ steps. However, the training time of the proposed adaptive batching method is approximately $4$ times and $2$ times shorter than that of the linear batching method and the no-batching method, respectively. In other words, the proposed method significantly reduces the training time for the IGL. In the remaining simulations, the trained IGL parameters are fixed as those from the last step of the proposed ES with the adaptive batching, without further updates.

\subsection{Online Graph Construction with DHF Bucketing Mode}
We evaluate the trained IGL NN using the online graph construction process designed in Section \ref{subsec:dhf_bucketing} with the bucketing mode of DHF when there are $K=1000$ STAs.

\begin{figure}[!t]
\centering
\includegraphics[scale=0.9]{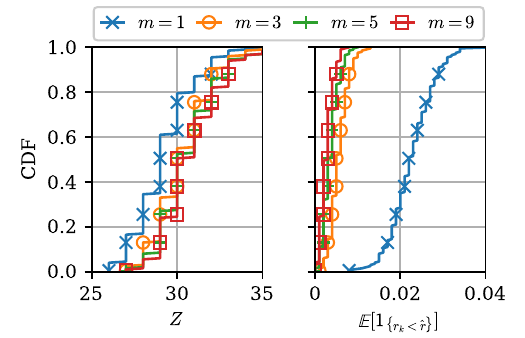}
\vspace{-0.2cm}
\caption{The IGL NN in the architecture when STAs are static.}
\label{fig:plot_static_fine_tune}
\vspace{-0.25cm}
\end{figure}
\begin{figure}[!t]
\centering
\includegraphics[scale=0.9]{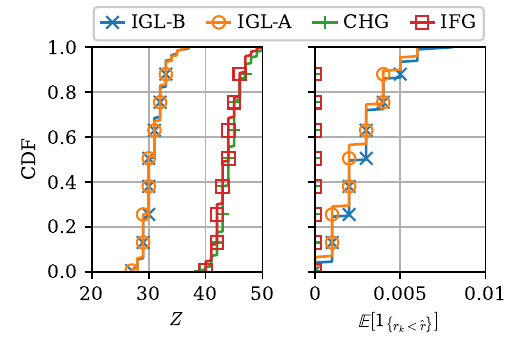}
\vspace{-0.2cm}
\caption{Comparison between the IGL and human intuition-based graphs.}
\label{fig:plot_static_fine_tune_cmp_graph}
\vspace{-0.25cm}
\end{figure}
\begin{figure}[!t]
\centering
\includegraphics[scale=0.9]{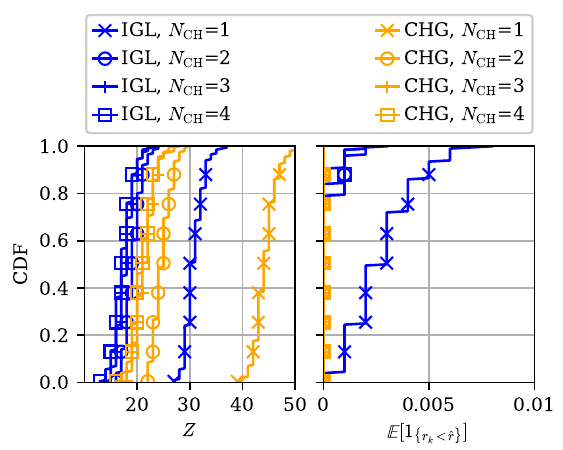}
\vspace{-0.2cm}
\caption{Performance when using different numbers $N_{\text{CH}}$ of channels.}
\label{fig:plot_static_fine_tune_channel_separation}
\vspace{-0.25cm}
\end{figure}

\subsubsection{Performance When STAs are Static}
Fig. \ref{fig:plot_static_fine_tune} shows the IGL NN's performance in the cumulative distribution function (CDF) of the number of used slots and the number of QoS violating STAs when the IGL NN operates for different numbers of rounds, $m$, in the architecture. In this architecture, all STA pairs connected by edges in previous rounds are combined into the processing STA pairs for the current round, as shown in \eqref{eq:online_sta_pair_augment}.
The results show that across different rounds, the number of slots to color the generated graph remains consistent, around $30$ slots. It also shows that the IGL NN in the first round results in $2\%$ of STAs experiencing QoS violations. However, by running the online process for more rounds, the QoS violations are fewer than $1\%$.
This improvement is due to the bucketing mode in each round, which can miss some contending and interfering STA pairs, as shown in previous simulations in Section \ref{subsec:simulation_dhf}. By adding the previously connected pairs in the IGL process as \eqref{eq:online_sta_pair_augment}, the likelihood of missing the processing of contending or interfering STA pairs is efficiently reduced.

Fig. \ref{fig:plot_static_fine_tune_cmp_graph} compares the number of slots and QoS violations achieved by the IGL with the DHF's bucketing mode when the number of rounds $m$ is 9 (with legend ``IGL-B''), the IGL with all STA pairs processed (with legend ``IGL-A''), and the human intuition-based graph models (with legends ``CHG'' and ``IFG'').
The results show that the IGL achieves $25\%$ fewer slots than the human intuition-based graph models. This is because the IGL allows slot sharing among STA pairs that are not heavily contending or interfering with each other, while the human intuition-based graph models separate all contending or interfering pairs. Consequently, the human intuition-based graph models achieve fewer QoS violations but use more slots or cause higher delays in periodicity.
The results also show that the scheme with partly processed STA pairs using the DHF's bucketing mode achieves performance similar to the one where the IGL NN processes all STA pairs. This implies that only a subset of STA pairs, e.g., those likely to be contending or interfering, require IGL NN processing, which further suggests that the DHF's bucketing mode reduces the complexity when constructing the graph.

In practice, Wi-Fi networks often operate on multiple channels to reduce contention and interference. Fig. \ref{fig:plot_static_fine_tune_channel_separation} shows the performance of the IGL NN with the DHF's bucketing mode when different numbers of channels, $N_{\text{CH}}$, are used. Here, we assume that each neighboring AP operates on a different channel when $N_{\text{CH}}>1$, and each channel has an equal bandwidth of $B=20$ MHz. The results show that using more channels reduces both the number of slots and the number of QoS violations. This is because more channels reduce contention and interference among STAs, allowing more STAs to share the same slot without causing QoS violations.
As shown in the results, the performance gain from adding channels eventually plateaus. Increasing the number of channels reduces inter-AP interference because neighboring APs can operate on different channels. Once channel separation among neighboring APs is sufficient, the remaining dominant interference comes mainly from STAs associated with the same AP. Therefore, adding more channels beyond this point provides little additional benefit.
We also compare the performance of the IGL with human intuition-based graph models under different numbers of channels. Here, we show only the CHG results, because IFG performs similarly to CHG, as shown previously. The results show that the IGL uses approximately $25\%$ fewer slots than CHG across different numbers of channels, while CHG yields fewer QoS violations than IGL. These results are consistent with the observations for $N_{\text{CH}}=1$.

\subsubsection{Performance When STAs are Mobile}
\begin{figure}[!t]
\centering
\includegraphics[scale=0.865]{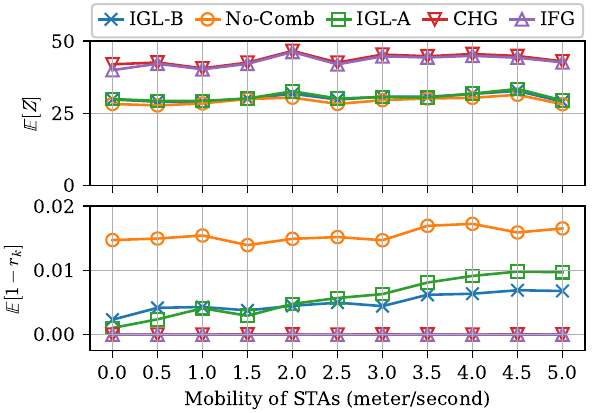}
\vspace{-0.5cm}
\caption{The IGL NN performance in online architecture with mobile STAs.}
\label{fig:plot_mobile_bler_nc}
\vspace{-0.25cm}
\end{figure}

We then show the IGL's performance when STAs are mobile. Specifically, STAs move in a random direction within the simulated area, with speeds varying from $0$ to $5$ meters per second, approximately the typical walking speeds of mobile robots. Note that when STAs reach the boundary of the simulated area, they move in another random direction toward the inside of the area.
We compare the following schemes: 1) the IGL using the DHF's bucketing mode and combining previous rounds' edges (the maximum rounds to combine is $I = 20$) into the processed pairs (with legend ``IGL-B''); 2) the IGL NN using the DHF's bucketing mode without combining previous edges in the processed pairs (with legend ``No-Comb''); and 3) the IGL NN processing all edge pairs among STAs (with legend ``IGL-A'').
Fig. \ref{fig:plot_mobile_bler_nc} measures the average number of slots and the average packet loss rates in the above schemes. The results show that the average number of slots is similar across all schemes (around $30$ slots), while the IGL-B and IGL-A schemes achieve lower packet loss rates. This is because the IGL-B and IGL-A schemes process more contending or hidden STA pairs than the No-Comb scheme. On the other hand, the IGL-B scheme has $30\%$ fewer packet losses than the IGL-A scheme when STAs have high mobility. This is because the DHF's bucketing mode selectively processes contending or interfering STA pairs, making the IGL process with the DHF's bucketing mode faster when constructing the graph. Consequently, it responds more timely to the varying locations of STAs. The detailed computing time of the IGL process is measured next. 
We also compare the performance of the IGL with human intuition-based graph models, CHG and IFG, when STAs are mobile in Fig. \ref{fig:plot_mobile_bler_nc}. The results show that the IGL-B scheme achieves $25\%$ fewer slots than CHG and IFG, while CHG and IFG achieve lower packet loss rates. This is because the human intuition-based graph models separate all contending or interfering STA pairs, while the IGL allows slot sharing among less contending or interfering STA pairs.

\subsection{Computational Complexity in Time}
\begin{figure}[!t]
\centering
\includegraphics[scale=0.875]{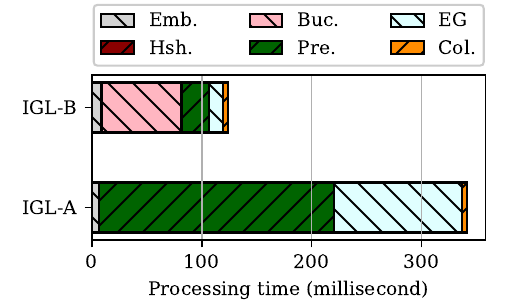}
\vspace{-0.2cm}
\caption{The computing time, $\expt[\tau^{(m)}]$, for the online graph constuction process to model and color the graph, including the time spent by the state embedding (Emb.), the DHF's hash code generation (Hsh.), the DHF's bucketing mode (Buc.), the predictors (Pre.), edge/graph generation (EG) and the graph coloring (Col.)}
\label{fig:plot_processing_time_profiling}
\vspace{-0.25cm}
\end{figure}

We discuss the computational complexity of the proposed methods in terms of the number of STAs, $K$. Training the state embedding (trained over STA-wise), the predictors (trained over STA-pairwise), and the DHF (also trained over STA-pairwise) incurs $\mathcal{O}(K)$, $\mathcal{O}(K^2)$, and $\mathcal{O}(K^2)$ complexity, respectively. The batching mode of the DHF requires a $\mathcal{O}(K)$ complexity to compare the random query and collect the batch. Training the EGNN in Algorithm \ref{alg:es-ggm} is also processed over STA-pairwise, resulting in an $\mathcal{O}(K^2)$ complexity.
The IGL complexity when constructing/processing the graph depends on the number of STA pairs collected using the bucketing mode of the DHF. 
Since it is difficult to analyze hash codes/tables generated by DHF, we instead measure the computing time of the online graph construction process quantitatively. 
Specifically, Fig. \ref{fig:plot_processing_time_profiling} compares the computing time of the online process when $K=1000$ in 1) the scheme where the IGL NN only processes STA pairs collected by the DHF bucketing mode (with legend ``IGL-B''), and 2) the scheme where the IGL NN processes all STA pairs (with legend ``IGL-A'').
The results show that the DHF's bucketing mode reduces the computing time for STA-pairwise operations, such as predictors and edge generation, by a factor of $8$, and the total computing time by a factor of $3$. This is because only the STA pairs in the buckets are processed, while the bucketing mode adds some additional computing time when collecting the STA pairs in the buckets/hash tables. Note that the current hash table construction is single-threaded on a central processing unit, while further time reduction can be achieved by an efficient parallelization on a graphics processing unit.

\begin{figure}[!t]
\centering
\includegraphics[scale=0.9]{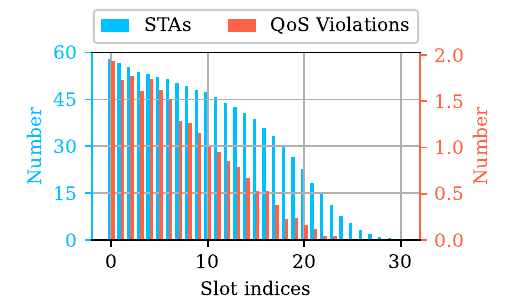}
\vspace{-0.2cm}
\caption{The STA and the QoS violation counts in different slots.}
\label{fig:plot_coloring_slot_idx}
\vspace{-0.25cm}
\end{figure}

\subsection{Impact of Greedy Graph Coloring}
Fig. \ref{fig:plot_coloring_slot_idx} shows the number of STAs and QoS violations in different slots. The results show that most STAs are assigned to slots with lower indices. This is because the greedy coloring scheme used in this work iteratively assigns STAs to previously added slots (with lower indices), while recently added slots remain almost empty. As a result, more QoS violations occur in the slots with more STAs. 
An efficient and balanced graph coloring scheme could further improve the QoS \cite{gu2025sigsdp}.

\section{Conclusion}\label{sec:conclusion}
This paper presented the IGL for managing contention and interference in Wi-Fi 7 networks through coloring-based slot assignment.
The proposed IGL framework achieved scalability in terms of network size in training and inference. Specifically, by employing the ES, the IGL eliminated the need for explicit STA-pairwise feedback, whose dimensionality grows quadratically with the number of STAs, enabling stable optimization even as the network scales. In addition, the DHF restricted training and inference to a small subset of likely contending or hidden STA pairs, significantly reducing the computational complexity in large-scale networks.
Extensive evaluation demonstrated the scalability and effectiveness of our methods in large Wi-Fi network scenarios.
For example, the proposed IGL framework with the DHF reduced the computing time of the online graph construction process by up to a factor of $3$, while achieving $25\%$ fewer slots than human intuition-based graph models with a slight increase in QoS violations.

Future work could focus on further optimizing the graph coloring algorithm and accelerating the computation routines in the IGL.
Moreover, the current IGL framework is designed as a centralized approach, where the AP collects the network states from all STAs and generates the interference graph for slot assignment. However, the proposed IGL framework can be extended to a distributed manner by enabling each STA to locally estimate its interference relationships with neighboring STAs based on local observations and limited information exchange. This can further reduce the communication overhead and improve scalability in large-scale networks.
Additionally, research could be conducted to explore richer graph representations that integrate the application layer's context, semantic, and task-oriented information, and to extend the framework to accommodate more heterogeneous network deployments, e.g., with devices with different traffic profiles (see Appendix D for detailed discussions).
Furthermore, the proposed IGL framework can be extended to the resource allocation schemes, including OFDMA and multi-user multiple-input multiple-output (MU-MIMO), in Wi-Fi 7 and beyond networks, by designing appropriate graph structures to represent the contention and interference in the schemes.
Finally, more efficient implementations of the graph coloring algorithms can be investigated to further balance the load among different slots and reduce the QoS violations, while exploiting the sparsity of the constructed graphs \cite{gu2025sigsdp}.

\bibliography{main}
\bibliographystyle{IEEEtran}

\begin{IEEEbiography}
[{\includegraphics[width=1in,height=1.25in,clip,keepaspectratio]{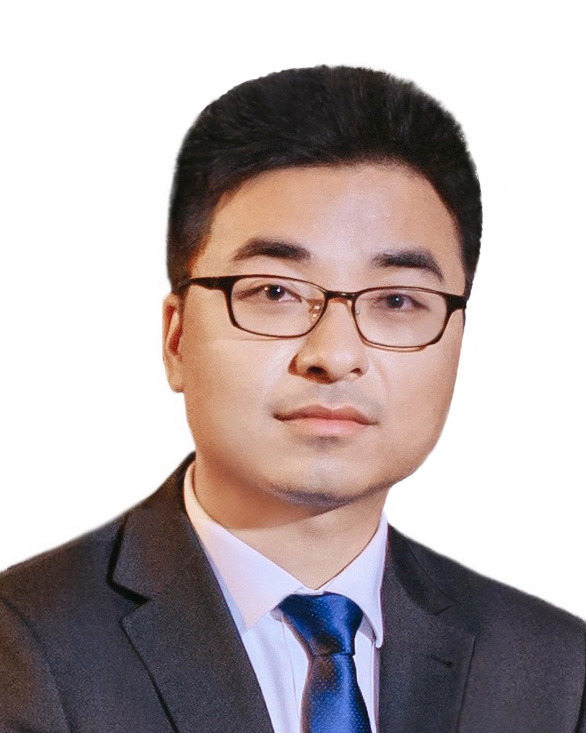}}]{Zhouyou Gu} completed his Ph.D. degree in 2023 at the School of Electrical and Information Engineering, the University of Sydney (USYD), Australia,
He was a research fellow at Deakin University, Australia in 2024. He is currently a research fellow at Singapore University of Technology and Design (SUTD), Singapore. 
He was a recipient of the Australian Research Training Program Stipend, as well as the USYD Postgraduate Research Supplementary and Completion Scholarships, the NVIDIA Academic Grant Program, and the Best Student Paper Award at ML4Wireless@AAAI 2026.
His research focuses on making wireless networks more intelligent at scale through graph learning, reinforcement learning, and AI-native protocol design.
\end{IEEEbiography}

\begin{IEEEbiography}
[{\includegraphics[width=1in,height=1.25in,clip,keepaspectratio]{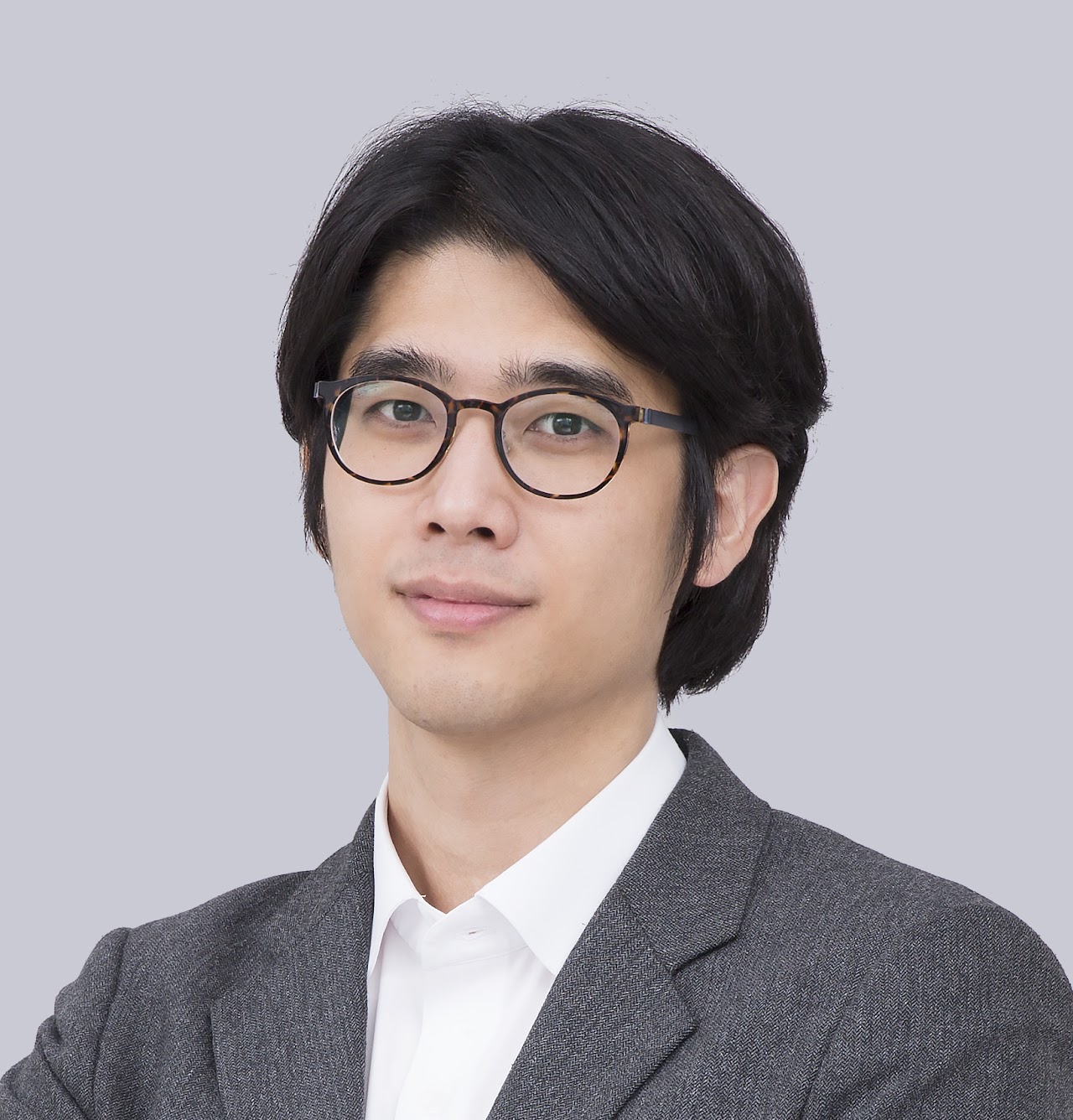}}]{Jihong Park} is an Associate Professor at the Singapore University of Technology and Design (SUTD) and an Honorary Associate Professor at Deakin University. He is Director of the MediaTek-SUTD Joint Laboratory and Deputy Director of the Future Communications Research and Development Programme (FCP) in Singapore. Dr. Park received his B.S. and Ph.D. degrees from Yonsei University, Seoul, South Korea, in 2009 and 2016, respectively. He was a Post-Doctoral Researcher with Aalborg University, Denmark, from 2016 to 2017, and the University of Oulu, Finland, from 2018 to 2019. His current research focus includes AI-native semantic communication and distributed machine learning for 6G. Dr. Park has served as Track Chair and Workshop Chair at leading conferences, including IEEE WCNC 2026, IEEE GLOBECOM 2023, and the 2025 ICML Workshop on Machine Learning for Wireless Communication and Networks. He has received several prestigious awards, including the 2023 IEEE Communication Society Heinrich Hertz Award and the 2022 FL-IJCAI Best Paper Award. Currently, Dr. Park is an Editor of IEEE Transactions on Communications and a Member of IEEE Signal Processing Society Machine Learning for Signal Processing Technical Committee, and Vice Chair of the AI-RAN Alliance’s AI-on-RAN Working Group.
\end{IEEEbiography}

\begin{IEEEbiography}
[{\includegraphics[width=1in,height=1.25in,clip,keepaspectratio]{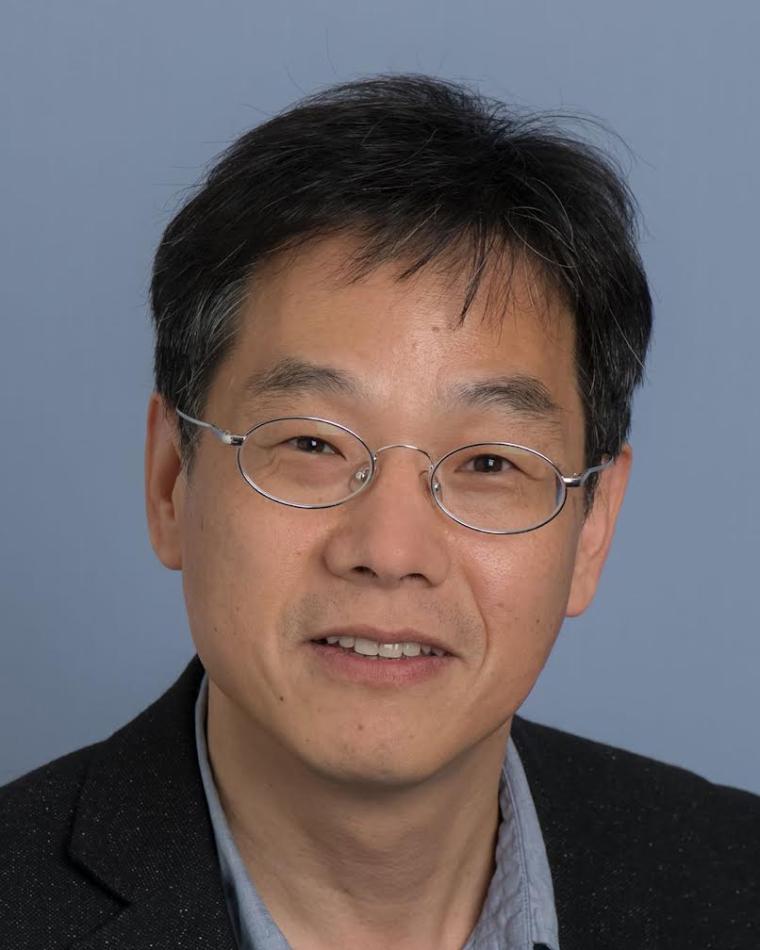}}]{Jinho Choi} was born in Seoul, Korea. He received B.E. (magna cum laude) degree in electronics engineering in 1989 from Sogang University, Seoul, and M.S.E. and Ph.D. degrees in electrical engineering from Korea Advanced Institute of Science and Technology (KAIST) in 1991 and 1994, respectively. He is with the School of Electrical and Mechanical Engineering, the University of Adelaide, Australia, as a Professor. His research interests include the Internet of Things (IoT), wireless communications, and statistical signal processing. He authored two books published by Cambridge University Press in 2006 and 2010 and one book by Wiley-IEEE in 2022. Prof. Choi received a number of best paper awards including the 1999 Best Paper Award for Signal Processing from EURASIP. He is a Fellow of the IEEE and has been on the list of World’s Top 2\% Scientists by Stanford University since 2020. Currently, he is a Senior Editor of IEEE Wireless Communications Letters and an Associate Editor of IEEE Trans. Mobile Computing. He has also served as a Division Editor of Journal of Communications and Networks (JCN), an Associate Editor or Editor of other journals including IEEE Trans. Communications, IEEE Communications Letters, JCN, IEEE Trans. Vehicular Technology, and ETRI journal.
\end{IEEEbiography}

\end{document}


\maketitle

\appendices

\section{Proof of Proposition \ref{prop:existence_of_optimal_binary_graph}}
\label{A:existence_of_optimal_binary_graph}
To prove this statement, we construct an example that satisfies it. Let $\hat{\mathbf{z}}$ and $\hat{Z}$ be an optimal solution of problem \eqref{eq:prob:cni_management_problem}. In the adjacency matrix $\hat{\mathbf{E}}$, we set $\hat{E}_{i,j} = 1$ for every STA pair $(i,j)$ with different colors and set $\hat{E}_{i,j} = 0$ otherwise. In other words, $\hat{\mathbf{E}}$ represents a complete multipartite graph.
The graph with $\hat{\mathbf{E}}$ has a unique coloring scheme $\hat{\mathbf{z}}$ and $\hat{Z}$ because any alternative coloring of STAs would cause violations of the edge constraints. Therefore, $\hat{\mathbf{z}}, \hat{Z} = \chi(\hat{\mathbf{E}})$, which also solves the contention and interference management problem according to the definition of $\hat{\mathbf{z}}$ and $\hat{Z}$.
By providing this example, we demonstrate that an optimal graph exists, thereby proving the statement.

\section{Proof of Theorem \ref{theorem:existence_of_optimal_binary_graph_based_on_CH}}
\label{A:existence_of_optimal_binary_graph_based_on_CH}
Given the optimal graph $\mathcal{G}^*$ with the adjacency matrix $\mathbf{E}^*$ and its coloring scheme $\mathbf{z}^*, Z^* = \chi(\mathbf{E}^*)$, the coloring scheme of the optimal graph, $\mathbf{z}^*$ and $Z^*$, corresponds to the optimal slot assignment, where all STAs' reliability constraints are satisfied, and the number of slots is minimized.
Suppose we remove all edges between non-contending and non-hidden STAs in $\mathcal{G}^*$. This results in a new graph $\mathcal{G}'$ with edge values $E'_{i,j} = E^*_{i,j} \land (O^{\mathrm{C}}_{i,j} + O^{\mathrm{H}}_{i,j})$ for all $i \neq j$. By re-coloring $\mathcal{G}'$, STAs in $\mathcal{G}'$ will share the same color (or the same slot) if and only if they have no edges between them in $\mathcal{G}'$, i.e., they are non-contending and non-hidden STAs, or they were in the same color in $\mathcal{G}^*$.
In this new color scheme, i.e., $\mathbf{z}', Z' = \chi(\mathbf{E}')$, each STA now shares the slot only with: 1) STAs that are neither contending with nor hidden from it, or 2) those STAs that were sharing slots in the previous optimal graph $\mathcal{G}^*$. Note that each STA's transmissions are not affected by non-contending and non-hidden STAs, and the slot-sharing STAs in the previous graph $\mathcal{G}^*$ have reliable transmissions, according to the definition of its optimality.
Therefore, the coloring scheme of the new graph $\mathcal{G}'$, i.e., $\mathbf{z}'$ and $Z'$, ensures reliable transmissions, thus satisfying the constraints. Note that $\mathcal{G}'$ is generated by edge removal from $\mathcal{G}^*$. Thus, their chromatic numbers satisfy $Z' \leq Z^*$.
Further, according to the optimality of $\mathcal{G}^*$, $Z^*$ is the minimum number of slots required to ensure reliability, implying that $Z'$ is equal to $Z^*$. Therefore, the coloring scheme of $\mathcal{G}'$, i.e., $\mathbf{z}'$ and $Z'$, ensures the reliability constraints and minimizes the number of slots. In other words, $\mathcal{G}'$ is optimal, which proves the statement.

\section{Pre-training of IGL NN}
\label{A:pre-training}

The SENN is trained as an encoder NN that maps the STA state sequence into an embedding vector, along with a decoder NN to recover the sequence from the embedding vector.
The decoder $\mathrm{DCNN}(\cdot|\theta^{\mu}_{\mathrm{DC}})$ has the same structure as the encoder but uses different parameters, $\theta^{\mu}_{\mathrm{DC}}$, and recovers the sequence from $\mathbf{v}_k$ as
\begin{equation}\label{eq:decoding_token}
\begin{aligned}
\tilde{\mathcal{S}}_k \triangleq  \{\tilde{\mathbf{s}}_k(1),\dots,\tilde{\mathbf{s}}_k(|\mathcal{A}_k|)\}  =  \mathrm{DCNN}(\mathbf{v}_k|\theta_{\mathrm{DC}}) , \ \forall k  ,
\end{aligned}
\notag
\end{equation}
where $\tilde{\mathcal{S}}_k$ is the recovered sequence. In the decoder, each LSTM block in the first layer takes $\mathbf{v}_k$ as input, and the recovered sequence is obtained as the collection of output gate vectors from the LSTM blocks in the last layer.
The loss function of the SENN is designed to minimize the difference between the original and recovered state sequences using the mean squared error as
\begin{equation}\label{eq:tokenizer_loss}
\begin{aligned}
L(\theta^{\mu}_{\mathrm{SE}} , \theta_{\mathrm{DC}}) = \frac{1}{K}\sum_{k=1}^{K}\sum_{a=1}^{|\mathcal{A}_k|} \|\mathbf{s}_k(a)  -\tilde{\mathbf{s}}_k(a) \|_2^2 .
\end{aligned}
\notag
\end{equation}
The SENN (the encoder NN) and the decoder NN are updated over the above loss functions using stochastic gradient descent with a learning rate $\eta$ for $M$ iterations.
Note that only the SENN (the encoder) is used in the IGL NN to embed the state sequences, as shown in Fig.~\ref{fig:ggm_structure}, while the decoder is not a component of the IGL NN.

The loss functions of the predictors are designed to minimize the difference between the true values and the predicted ones using binary cross-entropy as
\begin{equation}\label{eq:predictor_loss}
\begin{aligned}
L(\theta^{\mu}_{\mathrm{PC}}) &= \sum_{i\neq j}O^{\mathrm{C}}_{i,j}\log_2\tilde{O}^{\mathrm{C}}_{i,j} + (1-O^{\mathrm{C}}_{i,j})\log_2(1-\tilde{O}^{\mathrm{C}}_{i,j}) ; \\ 
L(\theta^{\mu}_{\mathrm{PH}}) &= \sum_{i\neq j}O^{\mathrm{H}}_{i,j}\log_2\tilde{O}^{\mathrm{H}}_{i,j} + (1-O^{\mathrm{H}}_{i,j})\log_2(1-\tilde{O}^{\mathrm{H}}_{i,j}) .
\end{aligned}
\notag
\end{equation}
The predictors are updated using stochastic gradient descent over the above loss functions with a learning rate $\eta$ for $M$ iterations.
Note that both predictor parameters, $\theta^{\mu}_{\mathrm{PC}}$ and $\theta^{\mu}_{\mathrm{PH}}$, are components of the IGL NN, as illustrated in Fig.~\ref{fig:ggm_structure}.

\section{Handling Heterogeneous STA Deployments}
\label{A:heterogeneous_STA_deployments}
For heterogeneous STA deployments with diverse traffic profiles, the IGL can be extended by augmenting each STA state $\mathcal{S}_k$ with traffic features, e.g., packet arrival rate, queue length, and priority level. This enriched representation allows the IGL NN to jointly model interference relations and traffic dynamics, leading to slot assignments that better balance reliability and traffic demand. The training set should therefore include diverse traffic patterns to improve robustness across heterogeneous scenarios. In addition, inter-STA traffic correlations (e.g., similar or complementary factory tasks) can be exploited: positive correlation generally increases co-activity and potential interference, whereas negative correlation reduces it. The IGL NN can learn these patterns to adjust edge values, and the DHF can incorporate the same traffic features in its hash input to cluster STAs with similar traffic behavior more effectively. These extensions enable the IGL framework to accommodate more complex and realistic network deployments, which will be studied in future work.
